\documentclass{ieee-ojvt}
\usepackage{amssymb}
\usepackage{cite}
\usepackage{amsmath,amsfonts}
\usepackage{algorithmic}
\usepackage{graphicx,color}
\usepackage{textcomp}
\usepackage{amsthm}
\usepackage{arydshln}
\usepackage{bbding}
\usepackage{bigints}
\usepackage{booktabs}
\usepackage[caption=false,font=normalsize,labelfont=sf,textfont=sf]{subfig}
\usepackage{diagbox}
\usepackage{epsfig,latexsym}
\usepackage{epstopdf}
\usepackage{algorithm}
\usepackage{float}
\usepackage{flushend}
\usepackage{indentfirst}
\usepackage{lastpage}
\usepackage{makecell}
\usepackage{mathtools}
\usepackage{multirow}
\usepackage{pifont}
\usepackage{psfrag}
\usepackage{setspace}
\usepackage{stfloats}
\usepackage{subfig}
\usepackage{subfloat}
\usepackage{float}
\usepackage{times}
\usepackage{verbatim}
\usepackage{url}
\usepackage{hyperref}
\usepackage{balance}
\usepackage{cleveref}
\usepackage[table]{xcolor} 
\usepackage{array}         

\theoremstyle{remark}
\newtheorem{theorem}{\quad \textbf{Theorem}}
\newtheorem{lemma}{\quad \textbf{Lemma}}

\newtheorem{remark}{\quad \textbf{Remark}}

\begin{document}

\title{Space-Air-Ground Integrated Networks: Their Channel Model and Performance Analysis}

\author{CHAO~ZHANG\authorrefmark{1} (Member, IEEE),
QINGCHAO LI\authorrefmark{1} (Member, IEEE), \\
CHAO XU\authorrefmark{1} (Senior Member, IEEE),
LIE-LIANG YANG\authorrefmark{1} (Fellow, IEEE),\\
AND LAJOS HANZO\authorrefmark{1} (Life Fellow, IEEE)}
\affil{School of Electronics and Computer Science, University of Southampton, Southampton SO17 1BJ, U.K.}
\corresp{CORRESPONDING AUTHOR: Lajos Hanzo (e-mail: lh@ecs.soton.ac.uk).}

\begin{abstract}
Given their extensive geographic coverage, low Earth orbit (LEO) satellites are envisioned to find their way into next-generation (6G) wireless communications. This paper explores space-air-ground integrated networks (SAGINs) leveraging LEOs to support terrestrial and non-terrestrial users. We first propose a practical satellite-ground channel model that incorporates five key aspects: 1) the small-scale fading characterized by the Shadowed-Rician distribution in terms of the Rician factor $K$, 2) the path loss effect of bending rays due to atmospheric refraction, 3) the molecular absorption modelled by the Beer-Lambert law, 4) the Doppler effects including the Earth's rotation, and 5) the impact of weather conditions according to the International Telecommunication Union Recommendations (ITU-R). Harnessing the proposed model, we analyze the long-term performance of the SAGIN considered. Explicitly, the closed-form expressions of both the outage probability and of the ergodic rates are derived. Additionally, the upper bounds of bit-error rates and of the Goodput are investigated. The numerical results yield the following insights: 1) The shadowing effect and the ratio between the line-of-sight and scattering components can be conveniently modeled by the factors of $K$ and $m$ in the proposed Shadowed-Rician small-scale fading model. 2) The atmospheric refraction has a modest effect on the path loss. 3) When calculating the transmission distance of waves, Earth's curvature and its geometric relationship with the satellites must be considered, particularly at small elevation angles. 3) High-frequency carriers suffer from substantial path loss, and 4) the Goodput metric is eminently suitable for characterizing the performance of different coding as well as modulation methods and of the estimation error of the Doppler effects.
\end{abstract}



\maketitle

\section{INTRODUCTION}

\IEEEPARstart{T}{he} next decade is expected to experience a proliferation of diverse wireless applications \cite{Zhu_Creating_2022_Aug,Xu_Sixty_2019}. Applications like navigation, underwater communication, and suburban communication demand extensive coverage to mitigate blind spots. However, terrestrial base stations (BSs) often face obstructions due to environmental factors, leading to unsatisfactory quality of service (QoS) \cite{Liu_Space_2018_4quarter}. As a potential alternative, satellites may establish line-of-sight (LoS) links with terrestrial receivers, thereby enhancing the coverage. Since Geostationary Earth Orbit satellites (GEOs) exhibit a latency approximately 100 times higher than that of Low Earth Orbit satellites (LEOs)~\cite{Ma_Satellite_2024_Feb,Su_Broadband_2019_Apr}, LEOs are better suited for delay-sensitive communications in space-air-ground integrated networks (SAGINs).

\begin{figure*}[t]
\centering
\includegraphics[width= 6.5in]{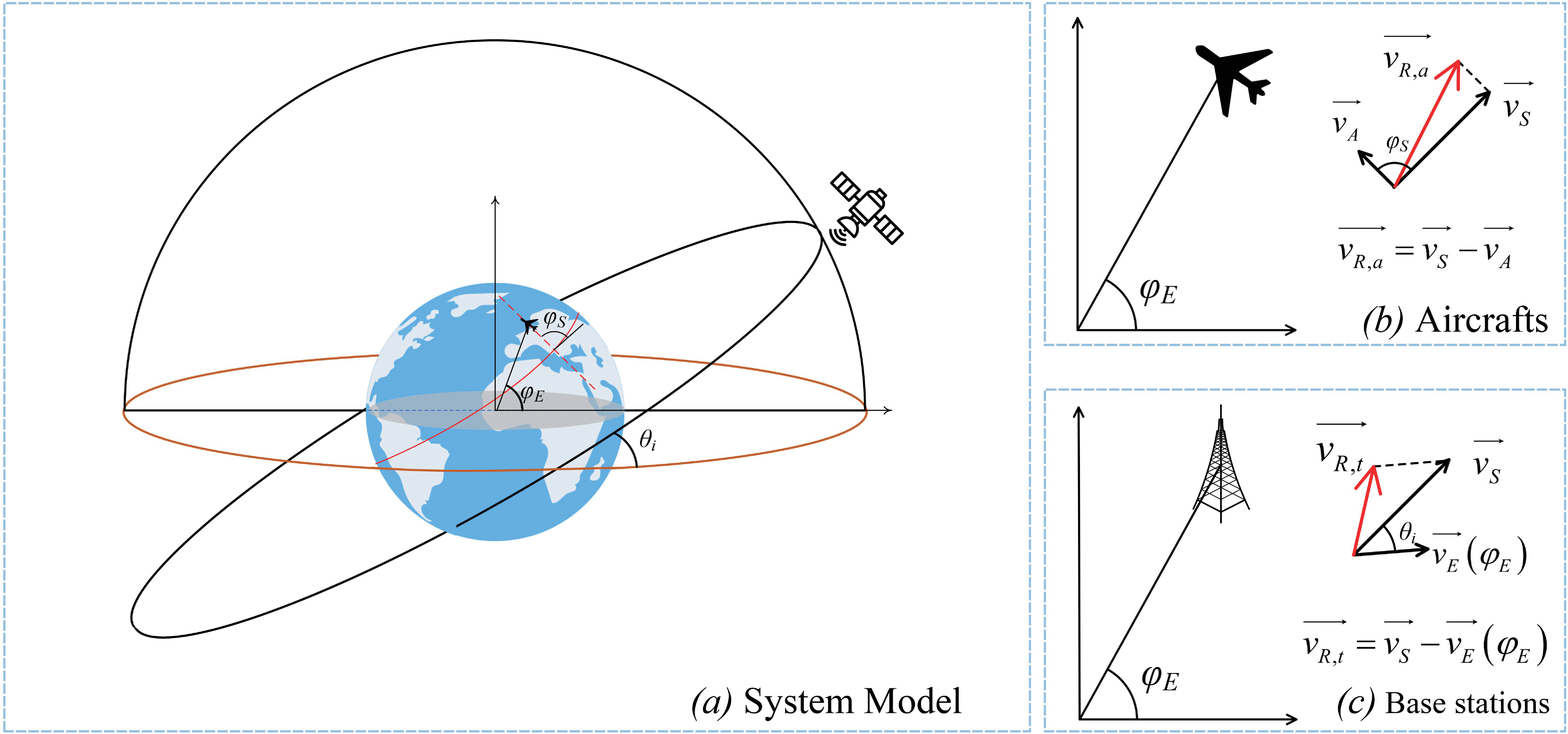}
\caption{Illustration of the system model: (a) The overview of the system model, (b) Aircrafts as users, and (c) Base stations or gateways as users.}
\label{systemmodel}
\end{figure*}

Naturally, new applications are always accompanied by novel challenges: 1) LEOs travel in space at a velocity of around 7.8 km per second with approximately 90 minutes to complete a full circle around the Earth, resulting in a severe Doppler shift and a short serving period. 2) Electromagnetic waves (the L band for the current satellite-ground communications) must propagate through the atmosphere from the LEOs to the ground, experiencing molecular absorption, especially in the Troposphere. The bands at higher frequencies than the L band (1-2 GHz) experience high absorption peaks owing to various gases. 3) The atmospheric turbulence imposes extra shadowing effects and multi-path propagation compared to terrestrial networks, which requires suitable small-scale fading models. 4) The refraction of the atmosphere results in bending rays, but their length must be shown to obtain the exact path loss. 5) The propagation of microwaves is influenced by the weather conditions as well, including rain, fog, and clouds.

To address these challenges, substantial research efforts have been dedicated to investigating SAGIN channel modeling and to their performance analysis, which are summarized as follows:
\begin{itemize}
\item \textbf{Small-scale fading}: A variety of distribution models have been considered for small-scale fading. Specifically, although the Log-Normal distribution fits the shadowing effects well~\cite{Hoang_Cloud_2024_May}, the Shadowed-Rician distribution~\cite{Abdi_Anew_2003_May} has mathematically-tractable expressions for fundamental channel statistics, which is widely harnessed for SAGINs. \cite{Pan_Performance_2020_Oct,An_Secure_2019_Nov,Ye_Earth_2021_Sep}. Although the Shadowed-Rician distribution is generally considered to be a subtype of Rician distribution, the analytical relationships between them, particularly regarding the distribution of the Rician factor $K$ and its properties in SAGINs, have yet to be fully explored.
\item \textbf{Path loss}: Again, due to the variation of atmospheric density, the transmission path of electromagnetic waves forms a bending ray. The authors of \cite{Bean_Models_1959_May,Blake_Ray_1968_Jan} have characterized the bending rays by the ray-tracing model. The additional distance caused by atmospheric refraction may lead to extra path loss, which has not been considered in the open literature. Additionally, the curvature of the Earth and the geometric relationship between the satellite and terrestrial users are often overlooked in the open literature.
\item \textbf{Molecular absorption}: In \cite{Hershberger_1946_Jun}, a gas absorption model has been proposed, which requires experimentally measured absorption coefficients. A theoretical physics-based model of molecular absorption has been constructed by the Born-Oppenheimer Approximation in \cite{Rosenkranz_Absorption_1993}. Based on diffusion loss, Sutherland and Bass~\cite{Sutherland_Atmospheric_2004_Feb} have derived absorption equations for various gases, including oxygen, nitrogen, carbon dioxide, and ozone. As a benefit of its mathematical tractability, the Beer-Lambert-law-based model has been widely exploited in Terahertz band airplane-satellite communications~\cite{Kokkoniemi_Channel_2021_Mar,Wang_Coverage_2023_Dec,Petrov_Interference_2017_Mar}. Despite the significant path loss encountered in Terahertz band communications, the molecular absorption model exploited might serve as a viable option in the context of SAGINs.
\item \textbf{Doppler effects}: Considering the geometric relationship between \a terrestrial user and a satellite, a normalized maximum Doppler frequency has been derived in \cite{Ali_Doppler_1998_Mar} along with its distribution in \cite{Papapetrou_Analytic_2005_Jul}. Given the normalized maximum Doppler frequency, the Doppler shift distribution has been modelled as a dubbed Jakes distribution in \cite{Haas_Aeronautical_2022_Mar}. Further mitigation of the Doppler effects has been achieved by non-coherent detection~\cite{Xu_Adaptive_2019_Feb}, Doppler frequency shift estimation~\cite{Zhang_Data_2020_Jan}, and orthogonal time frequency space (OTFS) modulation~\cite{Khammammetti_OTFS_2019,Gaudio_On_2020}. Nonetheless, the impact of the Earth's rotation on the relative velocity and Doppler effect have not been investigated.
\item \textbf{Weather conditions}: Weather conditions also influence the propagation of electromagnetic waves, which include rain \cite{ITU-R_rain,Zang_The_2019_Jun}, clouds \cite{Hoang_Cloud_2024_May,ITU-R_fog}, and fog \cite{ITU-R_fog}. Specifically, the International Telecommunication Union (ITU-R) recommends methods for evaluating the attenuation due to the weather conditions in decibels (dB) \cite{ITU-R_rain,ITU-R_fog}, but the attenuation in power-domain is not directly provided.
\end{itemize}
\begin{table*}[!htb]
\footnotesize
\begin{centering}
\caption{The Novelty Table of the Proposed Channel Model for Space-Air-Ground Integrated Networks.}
\label{Novelty_Table}
\begin{tabular}{|>{\raggedright\arraybackslash}p{1.8cm}
                |>{\raggedright\arraybackslash}p{1.5cm}
                |>{\raggedright\arraybackslash}p{1.5cm}
                |>{\raggedright\arraybackslash}p{1.5cm}
                |>{\raggedright\arraybackslash}p{1.5cm}
                |>{\raggedright\arraybackslash}p{1.5cm}
                |>{\raggedright\arraybackslash}p{1.5cm}
                |>{\raggedright\arraybackslash}p{1.5cm}
                |>{\raggedright\arraybackslash}p{1.5cm}|}
\hline
\multicolumn{9}{|c|}{\cellcolor[RGB]{255, 224, 189}The Novelty Table}\\[0.5ex]
\hline
\hline
\rowcolor{MyLightBlue!40}
Reference papers & Geometric Doppler frequency&Molecular absorption& Shadowed-Rician fading & The Earth's rotation & Weather conditions, inc. rain, fog, and clouds  & Distribution v.s. Rician $K$ factor & Geometric path loss with bending rays & Goodput analysis\\
\hline
\cellcolor{MyLightGreen!40} \cite{Ali_Doppler_1998_Mar,Papapetrou_Analytic_2005_Jul}&\cellcolor{MyLightRed!40}\checkmark &&&&&&&\\
\hline
\cellcolor{MyLightGreen!40} \cite{Hershberger_1946_Jun,Rosenkranz_Absorption_1993,Sutherland_Atmospheric_2004_Feb,Kokkoniemi_Channel_2021_Mar,Wang_Coverage_2023_Dec,Petrov_Interference_2017_Mar} & &\cellcolor{MyLightRed!40}\checkmark&&&&&&\\
\hline
\cellcolor{MyLightGreen!40}\cite{Abdi_Anew_2003_May,Pan_Performance_2020_Oct,An_Secure_2019_Nov,Ye_Earth_2021_Sep}& &&\cellcolor{MyLightRed!40}\checkmark&&&&&\\
\hline
\cellcolor{MyLightGreen!40}\cite{Ye_Earth_2021_Sep}& &&&\cellcolor{MyLightRed!40}\checkmark&&&&\\
\hline
\cellcolor{MyLightGreen!40}\cite{Hoang_Cloud_2024_May,ITU-R_rain,Zang_The_2019_Jun,ITU-R_fog}& &&&&\cellcolor{MyLightRed!40}\checkmark&&&\\
\hline
\rowcolor{MyLightRed!40}
\cellcolor{MyLightGreen!40}Our Model & \checkmark &\checkmark&\checkmark&\checkmark&\checkmark&\checkmark&\checkmark&\checkmark\\
\hline
\end{tabular}
\end{centering}
\end{table*}

Against this background, for the first time in literature, we propose an improved SAGIN channel model that incorporates the following practical considerations: 1) the normalized Doppler frequency along with the impact of the Earth's rotation, 2) the Beer-Lambert-law-based model of molecular absorption, 3) the Shadowed-Rician distribution in terms of the Rician factor $K$, 4) the path loss model considering geographic distances and bending rays, and 5) the effect of weather conditions in accordance with ITU-R's guidance. The novel contributions of this paper are contrasted to the state-of-the-arts in Table \ref{Novelty_Table}, which are further elaborated on as follows:
\begin{itemize}
 \item We propose a practical channel model for SAGINs, where LEOs are leveraged to directly serve SAGIN users. In this model, we consider two types of users: terrestrial and non-terrestrial users. Taking the Earth's rotation into account, we incorporate the Earth's angular velocity into the channel modeling between LEOs and terrestrial users. The Earth's rotation effect is also inherently embedded into the velocity of non-terrestrial users.
 \item We evaluate five key aspects of SAGIN channel modelling. Firstly, as for small-scale fading, we utilize the Shadowed-Rician channel and analyze the distribution in terms of the Rician factor $K$ that characterizes the LoS and non-line-of-sight (nLoS) components. Secondly, as for large-scale fading, we take into account the curvature of the Earth, the geographical relationship between users and satellites, and the atmospheric refraction. The length of the bending propagation trajectory is calculated by exploiting the ray-tracing model. We then derive the ``straight-line" distance between the user and the LEO as the benchmark for bending rays. We compare the satellites' true elevation angles and detected angle of arrival for electromagnetic waves as well. Subsequently, we present the Beer-Lambert-law-based model for molecular absorption, followed by the derivation of the normalized maximum Doppler frequency for both terrestrial and non-terrestrial users. Finally, we examine the effects of weather conditions in both dB and in terms of Watts (W).
  \item Based on the proposed SAGIN channel model, we have further quantify and analyse the following three pivotal performance metrics of SAGINs. Firstly, we calculate the achievable upper bound of bit-error-rates (BER) considering multiple quadrature amplitude modulation (M-QAM) methods.  Secondly, we derive the closed-form expression of outage probability (OP), which serves as an intermediate step for calculating the ergodic rates (ER). Thirdly, we derive the Goodput (GP) attained, representing the average value of flawless reception rates that the receiver acquires over an extended duration.
  \item Numerical results yield the following conclusions. 1) In the proposed Shadowed-Rician small-scale fading model of SAGINs, the parameters of $K$ and $m$ can be appropriately adjusted to model the line-of-sight/scattering conditions and the atmospheric shadowing effect, respectively. 2) For small elevation angles, factors such as the geographic distances, geometric angles, and the curvature of the Earth should be considered for accurate path loss estimation or modeling, by contrast, they may be neglected for large elevation angles. 3)  The atmospheric refraction model in ITU-R has a limited impact on path loss. 4) High-frequency carriers lead to significant path loss.
\end{itemize}

This paper is structured as follows. Section II introduces the SAGIN concept and derives the relative angular velocities between users and LEOs. Section III presents a practical channel model that examines the Doppler effect, the Earth's rotation, molecular absorption, the Shadowed-Rician distribution having the Rician factor $K$, bending rays, the Earth's curvature, and weather conditions. Section IV derives the closed-form expressions of BER, OP, ER, and GP. Section V provides our numerical results, followed by the conclusion in Section VI.

\section{SYSTEM MODEL}
Our SAGIN encompasses both terrestrial and non-terrestrial users communicating with a LEO. To mitigate interference, distinct sub-bands are utilized for the downlink and uplink channels. We assume that the Earth is a perfect sphere and the orbit of the LEO is a circle concentric with the Earth. Consequently, the angular velocities of both the Earth and of the LEO remain constant over time \footnote{We ignore the gravitation of both the moon and of other bodies in space.}. Based on these assumptions, the relative velocity between the LEO and the terrestrial or non-terrestrial user is derived.

\subsection{RELATIVE VELOCITY \& EARTH'S ROTATION}

Two categories of users are considered: airborne users (e.g., aircrafts and unmanned aerial vehicles~\cite{Xu_Adaptive_2019_Aug}) and ground users (e.g., BSs and gateways), as illustrated in Fig. \ref{systemmodel}. \emph{(b)} and Fig. \ref{systemmodel}. \emph{(c)}.

As for the non-terrestrial user, the Earth's rotation influences atmospheric turbulence, leading to time-varying velocities for different eastward/westward travelling directions. We define the non-terrestrial user's velocity as $\overrightarrow{{v_A}}$, which inherently incorporates the effects of the Earth's rotation. Consequently, the relative velocity between the LEO and the aircraft is given by:
\begin{align}
\left| {\overrightarrow {{v_{R,a}}} } \right| &{=} \left| {\overrightarrow {{v_S}}  - \overrightarrow {{v_A}} } \right| {=} \sqrt {{{\left| {\overrightarrow {{v_S}} } \right|}^2} + {{\left| {\overrightarrow {{v_A}} } \right|}^2} - 2{{\left| {\overrightarrow {{v_S}} } \right|}}\left| {\overrightarrow {{v_A}} } \right|\cos \left( {{\varphi _S}} \right)},
\end{align}
where ${\overrightarrow {{v_S}} }$ is the velocity of the LEO satellite and ${{\varphi _S}}$ is the angle between the direction of $\overrightarrow{{v_A}}$ and that of ${\overrightarrow {{v_S}} }$.

As for the terrestrial user, the effects of the Earth's rotation must be considered. The angular velocity of the Earth is denoted as ${\omega _E}$, and its radius is represented by $R$. Let us define the angle between the user and the equatorial plane as ${{\varphi _E}}$. Subsequently, we denote the angle between the orbit and the equatorial plane as $\theta_i$. The relative velocity between the satellite and the terrestrial user is derived as:
\begin{align}
& \left| {\overrightarrow {{v_{R,t}}} } \right| = \left| {\overrightarrow {{v_S}}  - \overrightarrow {{v_E}} \left( {{\varphi _E}} \right)} \right|\notag\\
 &= \sqrt {{{\left| {\overrightarrow {{v_S}} } \right|}^2} + {R^2}\omega _E^2{{\cos }^2}\left( {{\varphi _E}} \right) - 2\left| {\overrightarrow {{v_S}} } \right|R{\omega _E}\cos \left( {{\varphi _E}} \right)\cos \left( {{\theta _i}} \right)} .
\end{align}

Given the LEO's altitude, denoted as $H$, the relative angular velocity is given by
\begin{align}\label{a_v}
{\omega _{R,u}} = \frac{{\left| {\overrightarrow {{v_{R,u}}} } \right|}}{{R + H}},
\end{align}
where we have $u\in\{a,b\}$ ($u=a$ for the non-terrestrial users and $u=t$ for the terrestrial users). This relative angular velocity is exploited for calculating the normalized Doppler frequency, shown as \eqref{Doppler} in Section III. D.

\section{A PRACTICAL CHANNEL MODEL}

In satellite-ground communication, more severe challenges emerge than in terrestrial communication due to atmospheric interactions. Again, in our SAGIN model, we explicitly address the following five key factors: (1) small-scale fading due to atmospheric turbulence and shadowing, (2) ray bending caused by atmospheric refractivity, (3) molecular absorption of gases, (4) the Doppler effects, and (5) weather impacts, including rain, fog, and clouds.

\begin{theorem}\label{ChannelModel}
Considering the attenuation caused by the five factors, we model the received signal at the user side as:
\begin{align}
y = \sqrt {P_s{{\cal P}_{PL}}{{\cal P}_{abs}}{{\cal P}_{Rain}}{{\cal P}_{Fog}}{{\cal P}_{Clouds}}} {h_{ST}}x + n,
\end{align}
where $y$ is the received signal, $x$ is the transmitted signal, and $n$ is the additive white Gaussian noise (AWGN) whose variance is denoted as $\sigma^2$. We represent the transmit power from the LEO as $P_s$. The parameter $h_{ST}$ represents the small-scale fading. It obeys the Shadowed-Rician distribution, whose probability density function (PDF) and cumulative distribution function (CDF) are expressed as \eqref{PDF_h_ST_K} and \eqref{CDF_h_ST_K}, respectively. The path loss ${\cal P}_{PL}$ is formulated in \eqref{pathloss}. The attenuation of molecular absorption, denoted as ${\cal P}_{abs}$, is presented as \eqref{molecularabsorption}. The attenuation equations of fog, rain, and clouds (denoted as ${{\cal P}_{Fog}}$, ${{\cal P}_{Rain}}$, and ${{\cal P}_{Clouds}}$) are encapsulated in \eqref{rain} to \eqref{clouds}, respectively, in \textbf{\emph{Theorem} \ref{T_rain_fog_clouds}}.

Based on the formula of the received signal, the signal-to-noise ratio (SNR) at the user side is expressed as:
\begin{align}\label{SNR}
\gamma_{SNR} = \frac{P_s{{{\cal P}_{PL}}{{\cal P}_{abs}}{{\cal P}_{Rain}}{{\cal P}_{Fog}}{{\cal P}_{Clouds}}{{\left| {{h_{ST}}} \right|}^2}}}{{{\sigma ^2}}}.
\end{align}
\end{theorem}

\subsection{SMALL-SCALE FADING WITH SHADOWING EFFECT}

The scattering effects experienced in SAGINs due to the atmosphere differ from those in terrestrial scenarios. Besides substantial reflections from buildings or walls, the primary contributors to small-scale attenuation include atmospheric turbulence, shadowing caused by gases (such as water vapour and oxygen), and weather conditions (such as fog, clouds, and rain).

The Log-Normal distribution may effectively represent practical scenarios~\cite{Le_Could_2021_May}, but this model is characterized by complex expressions, leading to intractable derivations~\cite{Abdi_Anew_2003_May}. As a more convenient alternative, the Shadowed-Rician model has gained significant popularity~\cite{Pan_Performance_2020_Oct,An_Secure_2019_Nov}. It models the line-of-sight (LoS) component combined with atmospheric shadowing effects as a Nakagami-$m$ distribution. The scattering component is represented by a Rayleigh distribution~\cite{Abdi_Anew_2003_May}. The PDF of the Shadowed-Rician distribution is expressed as:
\begin{align}
{p_{{{\left| {{h_{ST}}} \right|}^2}}}(x) = &{\left( {\frac{{2{b_0}m}}{{2{b_0}m + \Omega }}} \right)^m}\frac{1}{{2{b_0}}}\exp \left( { - \frac{x}{{2{b_0}}}} \right)\notag\\
 &\times {}_1{F_1}\left( {m,1,\frac{{\Omega x}}{{2{b_0}\left( {2{b_0}m + \Omega } \right)}}} \right),
\end{align}
where $2{b_0}={\mathbb{E}[{P_{nLoS}}]}$ is the average received power of the nLoS component, while $\Omega = {\mathbb{E}[{P_{LoS}}]}$ is that of the LoS component. The confluent hypergeometric function of the first kind is denoted as ${}_1{F_1}(\cdot,\cdot;\cdot)$, and $m$ is the Nakagami-$m$ fading parameter. \footnote{Note that a Nakagami-$m$ distribution is a Log-Normal distribution when $m=\frac{1}{2}$ \cite{book_probability}, while we cannot fix $m=\frac{1}{2}$ but adjust $m$ to express the atmospheric shadowing effects.}

\begin{lemma}\label{Shadowed_Rician_channel}
The confluent hypergeometric function of the first kind can be transformed into a sum of exponential equations using Laguerre polynomials, thereby simplifying the derivations. Consequently, the PDF and CDF expressions are further derived for the Shadowed-Rician channel as:
\begin{align}\label{PDF_h_ST}
{f_{{{\left| {{h_{ST}}} \right|}^2}}}(x) {=} &{a_{ST}}\sum\limits_{k = 0}^{m - 1} { \binom{m-1}{k}\frac{{{{\left( {{c_{ST}}} \right)}^k}}}{{k!}}{x^k}\exp \left( { - {e_{ST}}x} \right)},\\
\label{CDF_h_ST}
{F_{{{\left| {{h_{ST}}} \right|}^2}}}(x) {=}& 1 {-} \sum\limits_{k = 0}^{m {-} 1} \binom{m {-} 1}{k} \sum\limits_{p = 0}^k {\frac{{\varsigma (k)}}{{p!}{\left( {{e_{ST}}x} \right)^{-p}}}} \exp \left( { {-} {e_{ST}}x} \right) ,
\end{align}
where we have the following parameters: ${a_{ST}} = \frac{1}{{2{b_0}}}{\left( {\frac{{2{b_0}m}}{{2{b_0}m + \Omega }}} \right)^m}$, ${b_{ST}} = \frac{1}{{2{b_0}}}$, ${c_{ST}} = \frac{\Omega }{{2{b_0}\left( {2{b_0}m + \Omega } \right)}}$, ${e_{ST}} = {b_{ST}} - {c_{ST}} = \frac{m}{{2{b_0}m + \Omega }}$, and $\varsigma (k) = {a_{ST}}\frac{{{{\left( {{c_{ST}}} \right)}^k}}}{{\left(e_{ST}\right)^{k + 1}}} = \frac{{{{\left( {2{b_0}m} \right)}^{m - k - 1}}{\Omega ^k}}}{{{{\left( {2{b_0}m + \Omega } \right)}^{m - 1}}}}$. Additionally, the binomial coefficient is denoted as $\binom{n}{k}=\frac{n!}{k!(n-k)!}$.
\begin{proof}
See Appendix~A.
\end{proof}
\end{lemma}

Subsequently, we define the Rician $K$ factor, denoted as $K = {\Omega}/{\left(2b_0\right)}$~\cite{Xu_OTFS_2023_Jan}, and normalize the mean value of the Shadowed-Rician variable as $2b_0+\Omega=1$\footnote{As calculated by \textbf{\emph{Lemma} \ref{L_E_h_ST}}, the expectation of the Shadowed-Rician distribution equals ``$2b_0+\Omega$".}. Consequently, the LoS and nLoS components are expressed as $\Omega =\frac{\Omega }{{2{b_0} + \Omega }} = \frac{K}{{K + 1}}$ and $2b_0=\frac{{2{b_0}}}{{2{b_0} + \Omega }} = \frac{1}{{K + 1}}$, respectively. Since the parameter $m$ of the Nakagami-$m$ distribution represents the shadowing effect of the LoS component, we define two parameters, including: 1) the shadowed LoS factor, denoted as ${K_{LoS}} = \frac{\Omega }{{\left( {2{b_0} + \Omega } \right)m}} = \frac{\Omega }{m} = \frac{K}{{\left( {K + 1} \right)m}}$, and 2) the isotropic scattering factor, denoted as ${K_{Sct}} = \frac{{2{b_0}}}{{2{b_0} + \Omega }} = 2{b_0} = \frac{1}{{K + 1}}$.

\begin{theorem}\label{D_K}
Given the normalized received power as $2b_0+\Omega=1$, the Shadowed-Rician distribution of the small-scale fading model, denoted as $K_{LoS} = \frac{K}{{\left( {K + 1} \right)m}}$ and $K_{Sct} = \frac{1}{{K + 1}}$, is expressed by the PDF and CDF expressions v.s. the Rician $K$ factor as:
\begin{align}\label{PDF_h_ST_K}
{f_{{{\left| {{h_{ST}}} \right|}^2}}}(x) = & \sum\limits_{k = 0}^{m - 1} \binom{m-1}{k}\frac{{{{\left( {{K_{Sct}}} \right)}^{m - k - 1}}{{\left( {{K_{LoS}}} \right)}^k}}}{{k!{{\left( {{K_{Sct}} + {K_{LoS}}} \right)}^m}}}\notag\\
&\hspace*{-1cm} \times {{\left( {\frac{x}{{{K_{Sct}} + {K_{LoS}}}}} \right)}^k}\exp \left( { - \frac{x}{{{K_{Sct}} + {K_{LoS}}}}} \right),
\end{align}
\begin{align}\label{CDF_h_ST_K}
{F_{{{\left| {{h_{ST}}} \right|}^2}}}(x) = & 1 - \underbrace {\sum\limits_{k = 0}^{m - 1} {\binom{m-1}{k}} \frac{{{{\left( {{K_{Sct}}} \right)}^{m - k - 1}}{{\left( {{K_{LoS}}} \right)}^k}}}{{{{\left( {{K_{Sct}} + {K_{LoS}}} \right)}^{m - 1}}}}}_{{\rm{binomial \hspace*{0.1cm} expansion  \hspace*{0.1cm} form}}}\notag\\
& \hspace*{-1cm}\times \sum\limits_{p = 0}^k {\frac{1}{{p!}}\underbrace {{{\left( {\frac{x}{{{K_{Sct}} + {K_{LoS}}}}} \right)}^p}\exp \left( { - \frac{x}{{{K_{Sct}} + {K_{LoS}}}}} \right)}_{{\rm{exponential  \hspace*{0.1cm} distribution  \hspace*{0.1cm} form}}}} ,
\end{align}
where ``$K_{LoS}+K_{Sct}$" represents all the components under the atmospheric shadowing effect with $K_{LoS}+K_{Sct} \le 1$. This clearly indicates that $K_{LoS}+K_{Sct} = 1$ if and only if $m=1$.
\end{theorem}

\begin{remark}
Since the parameter $m$ of the Nakagami-$m$ distribution only represents the atmospheric shadowing effect for the LoS component, the Shadowed-Rician distribution may create a misconception that the scattering component escapes from the atmospheric shadowing effect. Since the shadowed-Rician model constitutes a full-scattering scenario (isotropic reflection environment) for the nLoS component, the atmospheric shadowing effect has been considered. It is not necessary to introduce an additional parameter $m$ to represent the specific portion of shadowing effects within the scattering component. Quantifying the percentage of the scattering component due to the shadowing effect is inherently unfeasible.
\end{remark}

\begin{remark}
As for the shadowed LoS link, when $m=1$, the channel conditions are extremely hostile. As $m$ increases, the LoS link becomes more dominant, representing scenarios associated with weak atmospheric turbulence, clear skies without rain, fog, or clouds, and minimal terrestrial blockages.
\end{remark}

\subsection{LARGE-SCALE FADING FOR BENDING RAYS}

The operational satellite communications exploit the L-band (typically 1 GHz to 2 GHz), while higher-frequency carriers, such as millimetre-wave (mm-wave) carriers, are anticipated to dominate 6G communications due to the impending spectrum crunch. Consequently, the path loss may become significant. Atmospheric refraction, which results in longer distances compared to direct LoS links, may further exacerbate this situation. This subsection aims for investigating the impact of atmospheric refraction on the path loss in 6G SAGINs.

\begin{figure}[!htb]
\centering          
	\includegraphics[width= 3.4in]{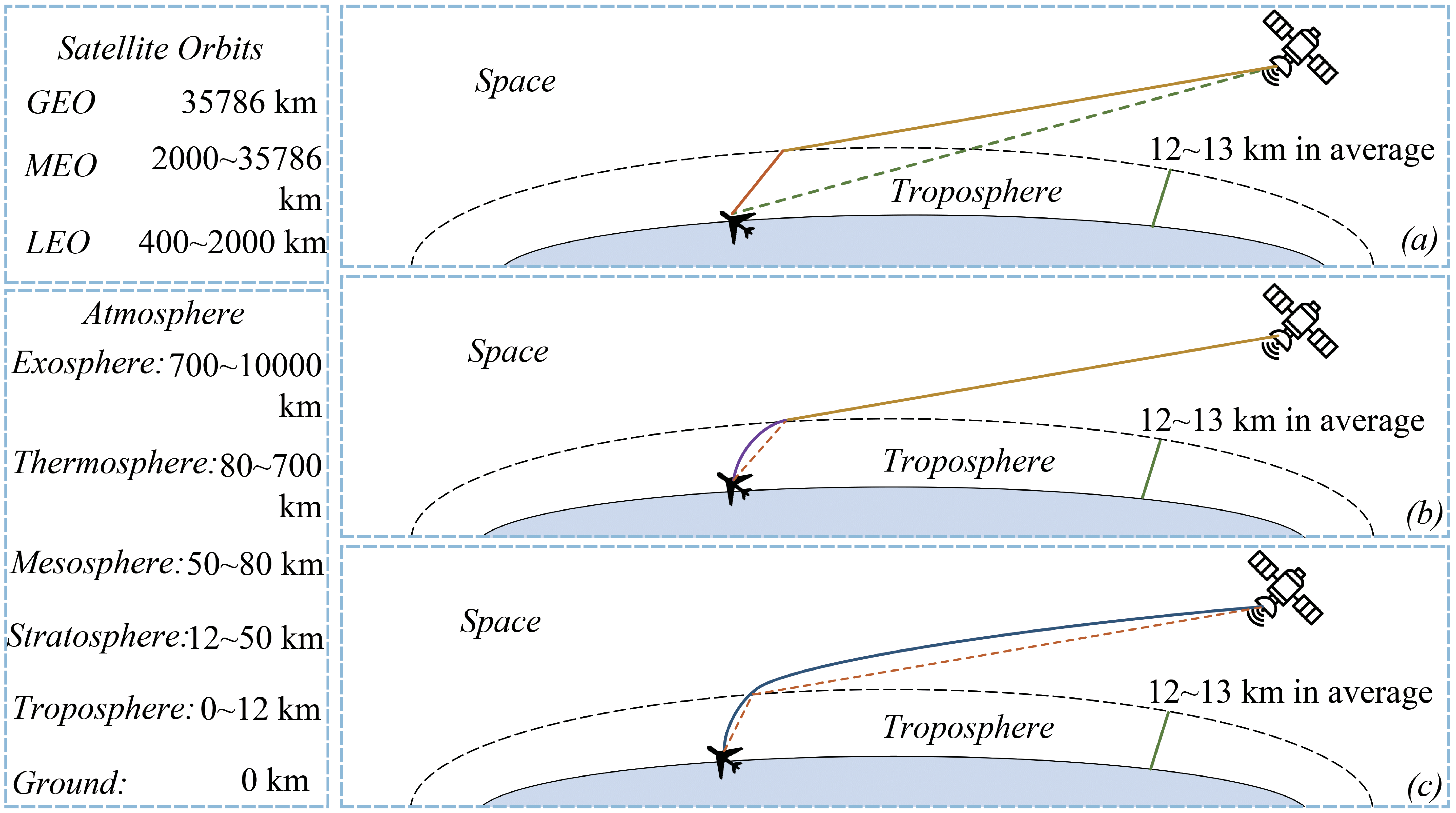}   
    \caption{Three models of bending rays for \textbf{\emph{Theorem} \ref{T_Bending}}.}
        \label{BendingRays}
\end{figure}

In Fig. \ref{BendingRays}, three models are depicted. (1) In Fig. \ref{BendingRays}.\emph{(a)}, a constant refractive index is assumed for both space and atmosphere, resulting in electromagnetic wave propagation mimicked by segmented broken lines. This scenario can be extended to multiple broken-line segments. (2) In Fig. \ref{BendingRays}.\emph{(b)}, wave propagation is straight in space but follows a curved trajectory in the atmosphere. (3) In Fig. \ref{BendingRays}.\emph{(c)}, the entire propagation path forms a continuous curve both in space and in the atmosphere.

Since the deviation of electromagnetic wave propagation is negligible in space above the atmosphere, the model in sub-figure \emph{(b)} fits practical scenarios well. Hence, it is chosen for evaluating the path loss in terms of the atmospheric refractive index formulated as $n = 1 + N \times 10^{-6}$ \cite{Bean_Models_1959_May}, where $N$ denotes the radio refractivity. The radio refractivity is influenced by atmospheric pressure, water vapour pressure, absolute temperature, and a number of other factors~\cite{ITU-R_Refraction}. Since horizontal variations in radio refractivity across adjacent regions are minimal, the refractive index may be modelled as a function of the vertical altitude above the Earth's surface. Based on the ITU-R recommendation sectors \cite{ITU-R_Refraction}, the refractive index as a function of the altitude is expressed as:
\begin{align}\label{refractivity}
n\left( h \right) =& 1 + {N_0} \times {10^{ - 6}} \times \exp \left( {-\frac{h}{{{h_0}}}} \right) \notag\\
= &1 + {N_0^{'}}\exp \left( {-\frac{h}{{{h_0}}}} \right),
\end{align}
where $N_0$ is the average value of the atmospheric refractivity extrapolated to sea level and $h_0$ is the altitude within the atmosphere (km).

\subsubsection{THE LENGTH OF THE BENDING PROPAGATION TRAJECTORY}
Three theorems are formulated in this subsection, namely \textbf{\emph{Theorem} \ref{T_Bending}} to \textbf{\emph{Theorem} \ref{T_straight}}. \textbf{\emph{Theorem} \ref{T_Bending}} characterizes the bending trajectory of electromagnetic waves in the face of atmospheric refractivity by a ray-tracing method. Subsequently, \textbf{\emph{Theorem} \ref{T_Bending2}} extends \textbf{\emph{Theorem} \ref{T_Bending}} with the objective of having higher accuracy. Finally, \textbf{\emph{Theorem} \ref{T_straight}} quantifies the straight length of  the link spanning from the user to the satellite used as a benchmark.

\begin{figure}[!htb]
\centering          
	\includegraphics[width= 3in]{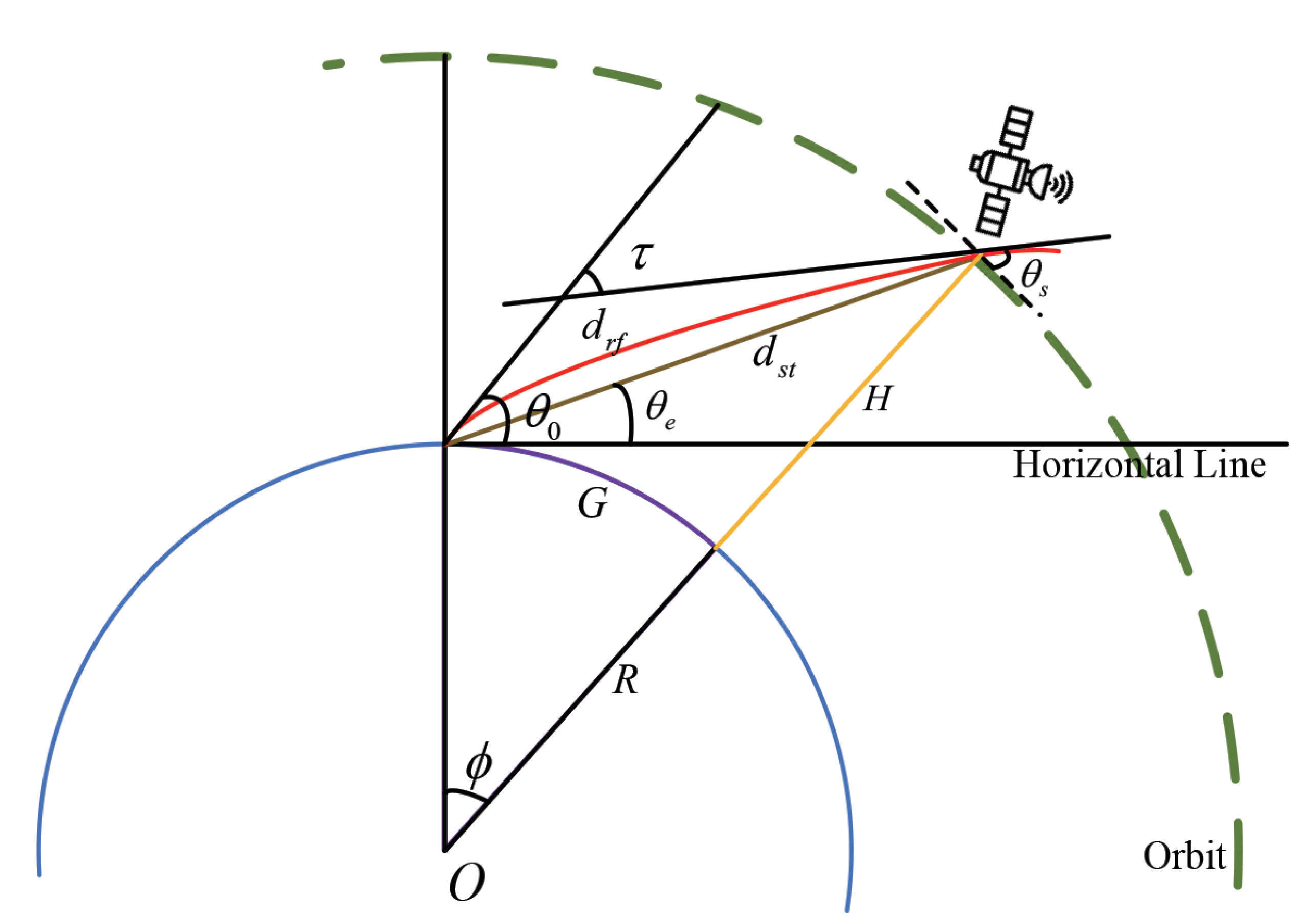}   
    \caption{The notation of angles and distances from the satellite to the ground user for \textbf{\emph{Theorem} \ref{T_Bending}}, \textbf{\emph{Theorem} \ref{T_Bending2}}, \textbf{\emph{Theorem} \ref{T_straight}}, and \textbf{\emph{Theorem} \ref{T_AOA}}.}
        \label{Angle_relationship}
\end{figure}

Before we present the following theorems, this paragraph defines parameters seen in Fig. \ref{Angle_relationship} for the convenience of reading. In Fig. \ref{Angle_relationship} explicitly, the blue semi-circular arc represents the Earth's surface along with its radius $R$ represented by the black solid line. The satellite orbits around the Earth along the green dashed line, where the orbit's altitude $H$ is indicated by the yellow solid line. The length $d_{rf}$ of the curved trajectory from the satellite to the ground is the red curve, whose tangent at the user side forms the angle $\theta_{0}$ with the horizontal line. The direct distance from the satellite to the user is $d_{st}$, which forms the true elevation angle $\theta_{e}$ with the horizontal line. The angle constituted by the tangent of the bending wave at the satellite side and that of the orbit is $\theta_{s}$. The two tangents of the bending wave (the red curve) both at the user side and at the satellite side intersect, forming an angle $\tau$. The length of the purple curve, denoted as $G$, represents the distance along the Earth's surface, which is defined as the ground range.

\begin{theorem}\label{T_Bending}
As shown in Fig. \ref{BendingRays}\emph{.(b)}, given the Earth's radius $R$, the altitude of the satellite's orbit $H$, and the initial (detected) elevation angle $\theta_0$, the length of the bending path is calculated as:
\begin{align}\label{d_rf_h}
{d_{rf}}\left( h \right) =& \sum\limits_{i = 1}^M {\frac{{H{\omega _i}}n\left( {{\kappa _i}} \right)}{2}{{\left( {1 - {{\cos }^2}\left( {\frac{2i-1}{{2M }}\pi } \right)} \right)}^{ \frac{1}{2}}}} \notag\\
&\times {\left( {1 - {{\left( {\frac{{{n_0}\cos \left( {{\theta _0}} \right)}}{{n\left( {{\kappa _i}} \right)\left( {1 + \frac{{{\kappa _i}}}{R}} \right)}}} \right)}^2}} \right)^{ - \frac{1}{2}}},
\end{align}
where we have ${\kappa _i} = \frac{{H\left( {\cos \left( {\frac{{2i - 1}}{{2M}}\pi } \right) + 1} \right)}}{2}$ and ${\omega _i} = \frac{\pi }{M}$. Recall that $n(\cdot)$ is shown as \eqref{refractivity}.
\begin{proof}
The differential equation of the curved trajectory's length is expressed as \cite{Blake_Ray_1968_Jan}:
\begin{align}\label{diff_bending}
{{dh} \mathord{\left/
 {\vphantom {{dh} {d{d_{rf}}}}} \right.
 \kern-\nulldelimiterspace} {d{d_{rf}}}} = \sin \left[ {{\theta _{\beta ,n}}\left( h \right)} \right].
\end{align}

Based on Snell's law \cite{Blake_Ray_1968_Jan,Thayer_ARapid_1967,Rayleigh_On_1893}, we can formulate an equation in terms of the local elevation angle, the satellite's altitude, and the refractive index, denoted as:
\begin{align}\label{Snell}
n\left( h \right)\cos \left[ {{\theta _{\beta ,n}}\left( h \right)} \right]\left( {R + h} \right) = {n_0}\cos \left[ {{\theta _{\beta ,n}}\left( 0 \right)} \right]R,
\end{align}
where ${{\theta _{\beta ,n}}\left( h \right)}$ is the elevation angle at the location above the Earth's surface with an altitude of $h$.

We arrive at ${n_0} = n\left( 0 \right) = 1 + {N_0} \times {10^{ - 6}}$ upon calculating the initial atmospheric refractivity on the ground based on \eqref{refractivity}. By substituting \eqref{refractivity} and \eqref{Snell} into \eqref{diff_bending}, the integral expression of ${d_{rf}}\left( h \right)$ may be expressed as:
\begin{align}
{d_{rf}}\left( h \right) = &\int_0^H {\frac{{n\left( h \right)dh}}{{\sqrt {1 - {{\cos }^2}\left[ {{\theta _{\beta ,n}}\left( h \right)} \right]} }}} \notag\\
 =& \int_0^H {\frac{{n\left( h \right)}}{{\sqrt {1 - {{\left( {\frac{{{n_0}\cos \left( {{\theta _0}} \right)}}{{n\left( h \right)\left( {1 + \frac{h}{R}} \right)}}} \right)}^2}} }}dh}.
\end{align}

With the aid of the Chebyshev-Gauss quadrature shown as Eq. [25.4.38] of \cite{mathbook}, the final expression \eqref{d_rf_h} of the bending path's length is derived\footnote{The definition of the Chebyshev-Gauss quadrature is expressed as $\int_{-1}^{+1} \frac{f(x)}{\sqrt{1-x^2}} d x \approx \sum_{i=1}^n w_i f\left(x_i\right)$ where we have $x_i=\cos \left(\frac{2 i-1}{2 n} \pi\right)$ and the weight $w_i=\frac{\pi}{n}$.}.
\end{proof}
\end{theorem}

\begin{theorem}\label{T_Bending2}
Based on the model shown in Fig. \ref{Angle_relationship}, the specific computational method given in \cite{Blake_Ray_1968_Jan} enhances the accuracy of \textbf{\emph{Theorem} \ref{T_Bending}}, particularly in the region around the vicinity of $h=0$ with a small $\theta_0$. The length of the bending trajectory of electromagnetic waves may be formulated as:
\begin{align}\label{d_rf_2}
{d_{rf}} = & \int_0^H {\frac{{{n^2}\left( h \right)\left( {1 + \frac{h}{R}} \right)}}{{\sqrt {\mu  + \upsilon \left( h \right) + \omega \left( h \right) + \upsilon \left( h \right)\omega \left( h \right)} }}} dh\notag\\
 = & \sum\limits_{i = 1}^M {\frac{{{\omega _i}H{n^2}\left( {{\kappa _i}} \right)\left( {1 + \frac{{{\kappa _i}}}{R}} \right)\sqrt {1 - {{\cos }^2}\left( {\frac{{2i - 1}}{{2M}}\pi } \right)} }}{{2\sqrt {\mu  + \upsilon \left( {{\kappa _i}} \right) + \omega \left( {{\kappa _i}} \right) + \upsilon \left( {{\kappa _i}} \right)\omega \left( {{\kappa _i}} \right)} }}} ,
\end{align}
and the ground range $G$ is estimated by setting $\frac{h}{R}$, yielding:
\begin{align}\label{G_computation2}
G =& \int_0^H {\frac{{\left( {1 + {\rho _0}} \right)\cos \left( {{\theta _0}} \right)dh}}{{\left( {1 + \frac{h}{R}} \right)\sqrt {\mu  + \upsilon \left( h \right) + \omega \left( h \right) + \upsilon \left( h \right)\omega \left( h \right)} }}} \notag \\
 =& \sum\limits_{i = 1}^M {\frac{{H{\omega _i}\left( {1 + {\rho _0}} \right)\cos \left( {{\theta _0}} \right)\sqrt {1 - {{\cos }^2}\left( {\frac{{2i - 1}}{{2M}}\pi } \right)} }}{{2\left( {1 + \frac{{{\kappa _i}}}{R}} \right)\sqrt {\mu  + \upsilon \left( {{\kappa _i}} \right) + \omega \left( {{\kappa _i}} \right) + \upsilon \left( {{\kappa _i}} \right)\omega \left( {{\kappa _i}} \right)} }}} ,
\end{align}
where the parameters and functions exploited are defined as:
\begin{subequations}
    \begin{align}
    \label{n_general}
    &n(x) = 1 + {\rho _0}\exp \left( { - kx} \right),\\
    &\mu  = {\left( {1 + {\rho _0}} \right)^2}{\sin ^2}{\theta _0} - 2{\rho _0} - \rho _0^2 ,\\
&\upsilon \left( {x} \right) = 2{\rho _0}\exp \left( { - kx} \right) + \rho _0^2\exp \left( { - 2kx} \right),\\
&\omega \left( {x} \right) = \frac{{2x}}{R} + \frac{{x^2}}{{{R^2}}},\\
&{\kappa _i} = \frac{{H\left( {\cos \left( {\frac{{2i - 1}}{{2M}}\pi } \right) + 1} \right)}}{2},\\
&{\omega _i} = \frac{\pi }{M}.
    \end{align}
\end{subequations}
\begin{proof}
The proof of derivations is in Appendix~B.
\end{proof}
\end{theorem}

\begin{remark}
Note that \textbf{\emph{Theorem} \ref{T_Bending2}} necessitates ${\theta _0} \ge 1.5^\circ$. If we have ${\theta _0} < 1.5^\circ$, negative values are obtained. This condition is satisfied in practical scenarios since LEO satellites generally work with higher elevation angles than $1.5^\circ$ for the sake of better QoS.
\end{remark}

\begin{remark}
\textbf{\emph{Theorem} \ref{T_Bending2}} provides an adjustable refractivity function, denoted as \eqref{n_general}. To match the settings of the refractivity function in ITU-R and \eqref{refractivity}, we opt for fixing the value of two parameters, denoted as $\rho_0 = N_0^{'}$ and $k=1/h_0$.
\end{remark}

\setcounter{equation}{25}
\begin{figure*}[hpb!]
\hrulefill
\begin{align}\label{Doppler}
\frac{{\Delta f}}{{{f_c}}} =  - \frac{1}{c}\frac{{R{H_{os}}\sin \left( {\psi \left( {t,{t_0}} \right)} \right)\cos \left( {{{\cos }^{ - 1}}\left( {\frac{{R\cos {\theta _{\max }}}}{{{H_{os}}}}} \right) - {\theta _{\max }}} \right){\omega _{R,u}}}}{{\sqrt {{R^2} + H_{os}^2 - 2R{H_{os}}\cos \left( {\psi \left( {t,{t_0}} \right)} \right)\cos \left( {{{\cos }^{ - 1}}\left( {\frac{{R\cos {\theta _{\max }}}}{{{H_{os}}}}} \right) - {\theta _{\max }}} \right)} }}.
\end{align}
\end{figure*}
\setcounter{equation}{18}

\begin{theorem}\label{T_straight}
We harness the same parameters and geometric relationships as in \textbf{\emph{Theorem} \ref{T_Bending2}}, which are shown in Fig. \ref{Angle_relationship}. By extending the results in \textbf{\emph{Theorem} \ref{T_Bending2}}~\cite{Bean_Models_1959_May}, the direct distance (the straight line $d_{st}$) from the satellite to the user is derived as:
\begin{align}\label{d_st}
{d_{st}} {=} \frac{{\left( {R + H} \right)\sin \phi }}{{\cos {\theta _e}}}{=} \sqrt { {{H^2} {+} 4R\left( {R {+} H} \right){{\sin }^2}\left( {\frac{G}{{2R}}} \right)} } ,
\end{align}
where we have $\phi  = \frac{G}{R}$ with the expression of $G$ as \eqref{G_computation2}.
\begin{proof}
This result is calculated by exploiting the law of sines in the triangle constituted by the satellite, the user, and the center of the Earth $O$. By exploiting that $\sin\left({\frac{\pi}{2}+\theta _e}\right) = \cos{\theta _e}$ and substituting \eqref{G_computation2} into \eqref{d_st}, we obtain the final result.
\end{proof}
\end{theorem}

\begin{remark}
The extra length of the wave's propagation due to the atmospheric refraction is expressed as:
\begin{align}
d_{\rm{dif}} = d_{rf}-d_{st}.
\end{align}
\end{remark}

\begin{remark}
The path loss is expressed as:
\begin{align}\label{pathloss}
{{\cal P}_{PL}} = {\left( {\frac{c}{{4\pi {f_c}}}} \right)^2}d_{rf}^{ - \alpha },
\end{align}
where $c$ is the speed of light, ${{f_c}}$ is the frequency of carrier, and $\alpha$ is the path loss exponent. To obtain the final path loss, we may exploit the results of $d_{rf}$ in \textbf{\emph{Theorem} \ref{T_Bending}} and \textbf{\emph{Theorem} \ref{T_Bending2}}.
\end{remark}

\begin{remark}
The length of the bending signal propagation trajectory (denoted as the $d_{rf}$ derived in \textbf{\emph{Theorem} \ref{T_Bending}} and \textbf{\emph{Theorem} \ref{T_Bending2}}) might be further exploited for satellite-aided integrated sensing and communications (ISaC) as well as for satellite-based cooperative localization and navigation networks.
\end{remark}

\subsubsection{THE TRUE ELEVATION ANGLE}

Upon extending the results in \textbf{\emph{Theorem} \ref{T_Bending2}} and \textbf{\emph{Theorem} \ref{T_straight}}~\cite{Blake_Ray_1968_Jan}, we arrive at a computational convenient technique of by calculating the true elevation angle, that is, the geometric angle of arrival (AOA), which is formulated in \textbf{\emph{Theorem} \ref{T_AOA}}.

\begin{theorem}\label{T_AOA}
Upon harnessing the result of $G$ in \textbf{\emph{Theorem} \ref{T_Bending2}} and that of $d_{st}$ in \textbf{\emph{Theorem} \ref{T_straight}}, the true elevation angle is expressed as:
\begin{subequations}
\begin{align}
{\theta _e} &= \arcsin \left( {\frac{H}{{{d_{st}}}} + \frac{{{H^2}}}{{2R{d_{st}}}} - \frac{{{d_{st}}}}{{2R}}} \right)\left( {{\theta _0} \le \frac{\pi }{4}} \right),\\
{\theta _e} &= \frac{\pi }{2} - \arcsin \left( {\frac{{\left( {R + H} \right)\sin \left( {\frac{G}{R}} \right)}}{{{d_{st}}}}} \right)\left( {{\theta _0} > \frac{\pi }{4}} \right).
\end{align}
\end{subequations}
\end{theorem}

\begin{remark}
The results of the true elevation angle might be exploited for investigating the effect of the polarization of antenna arrays and the LEO-aided sensing networks, including localization and navigation.
\end{remark}

\begin{remark}
The elevation angle is modified by $\left({\theta _0} - \theta_e \right)$ radians due to the atmospheric refraction.
\end{remark}

\subsection{MOLECULAR ABSORPTION}

When the energy of electromagnetic waves at specific frequencies matches the energy required by free electrons in molecules to be transited from lower to higher energy states, substantial absorption occurs. In the atmosphere, key absorbing molecules include water vapor, oxygen, nitrogen, and their isotopes. Notably, the exploitation of high-frequency bands will result in numerous molecular absorption peaks, significantly degrading wireless communication performance~\cite{Petrov_Interference_2017_Mar}.

By combining both tractability and generality, the Beer-Lambert-law-based model has been formulated as \cite{Kokkoniemi_Channel_2021_Mar,Wang_Coverage_2023_Dec}:
\begin{align}\label{molecularabsorption}
{{\cal P}_{abs}}\left( {{f_{abs}},{r_{abs}}} \right) = {\exp{\left( \sum\nolimits_i {\kappa _{abs}^i\left( {{f_{abs}}} \right){r_{abs}}} \right)}},
\end{align}
where the parameters include: 1) $r_{abs}$ is the thickness of the medium, and 2) ${\kappa _{abs}^i\left( {{f_{abs}}} \right)}$ is the absorption coefficient of the $i^{th}$ absorbing species (molecules or their isotopologue) at the frequency of $f_{abs}$. We would have a fixed $r_{abs}$ for homogeneous medium and a variable $r_{abs}$ for non-homogeneous medium \cite{Kokkoniemi_Channel_2021_Mar}.

\subsection{THE DOPPLER EFFECT}

Due to the Doppler effect, the observed frequency and the change of frequency are generally calculated as a function of the relative velocity, expressed as:
\begin{align}
f_{ob} =& \frac{c \pm v_o}{c \pm v_s}f_{s}\approx \left(1+\frac{\Delta v}{c}\right)f_c,\\
\Delta f = & f_{ob} - f_{s} \approx \frac{\Delta v}{c} f_{s},
\end{align}
where again, $c$ is the speed of light. Furthermore, $f_{ob}$ and $v_o$ are the frequency and the velocity of the observer, respectively. Additionally, the frequency and the velocity of the source is respectively expressed as $f_s$ and $v_s$. We have ${\Delta v}$ as the relative velocity between the source and the target, denoted as $\left| {{v_o} - {v_s}} \right|$. Note that the frequency of the source is that of the carrier wave, i.e., $f_c=f_s$.

\begin{theorem}\label{T_Doppler}
When we consider both the geographic locations of users, the orbits of LEOs, and the rotation of the Earth, the Doppler effect is also influenced by the relative position between the users and the satellites \cite{Ali_Doppler_1998_Mar}. Naturally, it is impossible that each user has a satellite passing directly overhead. Explicitly, considering a maximum elevation angle between the user and the satellite is more practical. The derivation of the normalized Doppler frequency is expressed as \eqref{Doppler}, where ${{\omega _{R,u}}}$ is the relative angular velocity for the satellite, derived as \eqref{a_v}, and $\theta_{max}$ is the maximum elevation angle. Additionally, we have $\dot \psi (t) = \frac{{d\psi (t)}}{{dt}} = {\omega _{R,u}}$, $\psi \left( {t,{t_0}} \right) = \psi (t) - \psi \left( {{t_0}} \right)$.
\setcounter{equation}{26}
\begin{proof}
Proved in Appendix~D.
\end{proof}
\end{theorem}

\subsection{WEATHER CONDITIONS}

Based on the recommendations of the ITU, the attenuation due to rain, fog, and clouds has been evaluated by considering practical data sets \cite{ITU-R_rain,ITU-R_fog}. As for the rain model, the frequencies are in the range spanning from $1$ to $1000$ GHz. Additionally, the attenuation model for fog and clouds is valid for frequencies below $200$ GHz, which is based on the Rayleigh scattering. The attenuation in dB is expressed in \textbf{\emph{Lemma} {\ref{L_rain_fog_clouds}}} and the attenuation harnessed in our SNR expressions are presented in \textbf{\emph{Theorem} {\ref{T_rain_fog_clouds}}}.

\begin{lemma}\label{L_rain_fog_clouds}
The attenuation model for rain with the unit of dB/km is expressed as:
\begin{align}\label{rain_dB}
\gamma_R=K_R R_{rate}^{\alpha_R},
\end{align}
where $K_R$ and $\alpha_R$ are determined as functions of the carrier frequency, which are found in \cite{ITU-R_rain}, while $R_{rate}$ is the rain intensity (mm/h).

Additionally, the attenuation model for fog with the unit of dB/km is expressed as:
\begin{align}\label{fog_dB}
\gamma_F=K_L M_{den},
\end{align}
where $K_L$ is the specific attenuation coefficient in terms of liquid water density in the cloud or fog, denoted as $M_{den}$ \cite{ITU-R_fog}.

Moreover, the attenuation model for clouds with the unit of dB is expressed as:
\begin{align}\label{clouds_dB}
\gamma_C = \frac{L_W K_L}{\sin{\theta_e}} ,
\end{align}
where $L_W$ is the statistics of the total columnar content of liquid water and $\theta_e$ is the true elevation angle \cite{ITU-R_fog}.
\end{lemma}

\begin{theorem}\label{T_rain_fog_clouds}
Instead of considering the accumulated attenuation v.s. the distance instead of in dB, we consider the attenuation as a multiplicative factor in the SNR expression. Hence, the specific attenuation formulas for rain, fog, and clouds are expressed as follows.

As for the rain, the attenuation model is expressed as:
\begin{align}\label{rain}
{{\cal P}_{Rain}} ={10^{ - \frac{{{\gamma _R}{d_{Rain}}}}{{10}}}},
\end{align}
with $d_{Rain}$ as the length of the path that the electromagnetic waves go through in the rainfall area.

As for the fog, the attenuation model is expressed as:
\begin{align}\label{fog}
{{\cal P}_{Fog}} = 10^{-\frac{\gamma_F d_{Fog}}{10}},
\end{align}
with $d_{Fog}$ as the length of the path that the electromagnetic waves propagate through in the foggy area.

As for the clouds, the attenuation model is expressed as:
\begin{align}\label{clouds}
{{\cal P}_{Clouds}} = \left( {\prod\limits_{i = 1}^I {{{10}^{\frac{{{\gamma _{C,i}}}}{{10}}}}}}\right)^{-1} ,
\end{align}
with $\gamma_{C,i}$ as the attenuation for each cloud and the number of clouds is $I$.
\begin{proof}
As for the rain, we have a total rain's attenuation expressed as $\left({\gamma _R}{d_{Rain}}\right)$ dB. We have ${\gamma _R} = 10\times{\log _{10}}\left( {P_1/P_2} \right)$, where $P_1$ and $P_2$ are the powers of waves before and after the effect of rain (per km), respectively. Hence, we have ${P_{receiv}} = {10^{\left( {\frac{{{\gamma _R}{d_{Rain}}}}{{10}}} \right)}}{P_{trans}} = {{\cal P}_{Rain}}{P_{trans}}$. Consequently, the attenuation of rain, denoted as ${\cal P}_{Rain}$, is obtained. The attenuation expressions of other weather conditions may be similarly derived.
\end{proof}
\end{theorem}

\section{PERFORMANCE ANALYSIS: THE GOODPUT}

Based on the practical channel model presented in the above section, in this section, the GP performance of SAGINs is investigated. As we consider the long-term performance, we assume that the Doppler effect is averaged out. The impact of attenuation will be further evaluated by the SNR expression formulated in \eqref{SNR}.

\subsection{BIT-ERROR-RATE}

A BER bound has been evaluated in references \cite{Goldsmith_Variable_1997_Oct,Jiho_Transmit_2003_Feb,Foschini_Digital_1983_Feb} for an AWGN channel relying on M-QAM and perfect coherent phase detection:
\begin{align}
{\rm{BER}} \le 2\exp \left( {\frac{{ - 3{\mathbb{E}}\left[ {{\gamma _{SNR}}} \right]}}{{2\left( {M - 1} \right)}}} \right),
\end{align}
where $\mathbb{E} \left[\cdot\right]$ is the expectation.

Under the assumption of $M \ge  4$ and $0 \le {\mathbb{E}}\left[ {{\gamma _{SNR}}} \right] \le 30 $ dB, we have a tighter upper bound expressed as:
\begin{align}\label{BER_bound}
{\rm{BER}}  \le \frac{1}{5}\exp \left( {\frac{{ - 3{\mathbb{E}}\left[ {{\gamma _{SNR}}} \right]}}{{2\left( {M - 1} \right)}}} \right).
\end{align}
In the following, we derive the $\mathbb{E}\left[ {{\gamma _{SNR}}} \right]$ to obtain final BER bound.

\begin{lemma}\label{L_E_h_ST}
By exploiting the PDF of the Shadowed-Rician fading channel in \eqref{PDF_h_ST}, the expectation of ${{{\left| {{h_{ST}}} \right|}^2}}$ is derived as:
\begin{align}
 \mathbb{E}\left[ {{{\left| {{h_{ST}}} \right|}^2}} \right] = 2{b_0} + \Omega.
\end{align}
\begin{proof}
The proof is provided in Appendix~E.
\end{proof}
\end{lemma}

\begin{remark}
The normalization $2{b_0} + \Omega =1$ means that the expectation of $ {{{\left| {{h_{ST}}} \right|}^2}} $ equals $1$. This echoes the parameter settings of \textbf{\emph{Theorem} \ref{D_K}}.
\end{remark}

\begin{lemma}\label{L_E_SNR}
Recall that we normalize the expectation of $\mathbb{E}\left[ {{{\left| {{h_{ST}}} \right|}^2}}\right]$ as $2{b_0} + \Omega=1$. Given the SNR expression in \eqref{SNR}, we now formulate the expectation of the SNR as:
 \begin{align}
 {\mathbb{E}}\left[ {{\gamma _{SNR}}} \right] = {\mathbb{E}}\left[ {{\lambda _t}{{\left| {{h_{ST}}} \right|}^2}} \right] = {\lambda _t}{\mathbb{E}}\left[ {{{\left| {{h_{ST}}} \right|}^2}} \right] = {\lambda _t},
 \end{align}
where we have ${\lambda _t} = \frac{{{P_s}{P_{PL}}{P_{abs}}{P_{Rain}}{P_{Fog}}{P_{Clouds}}}}{{{\sigma ^2}}}$.
\end{lemma}

\begin{theorem}\label{T_BER}
With the aid of \textbf{\emph{Lemma} \ref{L_E_SNR}} and \eqref{BER_bound}, the upper bound of the average long-term BER under the M-QAM is expressed as:
 \begin{align}\label{BER_final}
 {\rm{BER}} \le \frac{1}{5}\exp \left( {\frac{{ - 3{\lambda _t}}}{{2\left( {M - 1} \right)}}} \right).
 \end{align}
\end{theorem}

\subsection{OUTAGE PROBABILITY}

This subsection calculates the outage probability (OP), which is a popular performance metric also required for calculating the ergodic rates (ER). We first derive the CDF of $\gamma_{SNR}$ by \textbf{\emph{Lemma} \ref{L_CDF_SNR}}, followed by the derivation of the OP.

\begin{lemma}\label{L_CDF_SNR}
With the aid of the Rician distribution in \textbf{\emph{Theorem} \ref{D_K}}, the CDF expression of $\gamma_{SNR}$ is derived as:
\begin{align}
{F_{{\gamma _{SNR}}}}(x) = & 1 - \sum\limits_{k = 0}^{m - 1} {\binom{m-1}{k}\frac{{{{\left( {{K_{Sct}}} \right)}^{m - k - 1}}{{\left( {{K_{LoS}}} \right)}^k}}}{{{{\left( {{K_{Sct}} + {K_{LoS}}} \right)}^{m - 1}}}}} \notag\\
&\times \sum\limits_{p = 0}^k {\frac{1}{{p!}}} {\left( {\frac{x}{{{\lambda _t}\left( {{K_{Sct}} + {K_{LoS}}} \right)}}} \right)^p} \notag\\
&\times \exp \left( { - \frac{x}{{{\lambda _t}\left( {{K_{Sct}} + {K_{LoS}}} \right)}}} \right).
\end{align}
\begin{proof}
This lemma is proved by substituting the SNR expression into \eqref{CDF_h_ST_K}, leading to the following result:
\begin{align}
{F_{{\gamma _{SNR}}}}(x) =& \Pr \left\{ {{\lambda _t}{{\left| {{h_{ST}}} \right|}^2} < x} \right\}={F_{{{\left| {{h_{ST}}} \right|}^2}}}\left( {\frac{x}{{{\lambda _t}}}} \right).
\end{align}
\end{proof}
\end{lemma}

\begin{theorem}\label{T_OP}
Given the outage threshold, denoted as $\gamma_{th}$, and the derivations in \textbf{\emph{Lemma} \ref{L_CDF_SNR}}, the OP of this SAGIN is expressed as:
 \begin{align}
{P_{out}} = & {F_{{\gamma _{SNR}}}}\left( {{\gamma _{th}}} \right)\notag\\
 =& 1 - \sum\limits_{k = 0}^{m - 1} {\binom{m-1}{k}\frac{{{{\left( {{K_{Sct}}} \right)}^{m - k - 1}}{{\left( {{K_{LoS}}} \right)}^k}}}{{{{\left( {{K_{Sct}} + {K_{LoS}}} \right)}^{m - 1}}}}} \notag\\
&\times \sum\limits_{p = 0}^k {\frac{1}{{p!}}} {\left( {\frac{{{\gamma _{th}}}}{{{\lambda _t}\left( {{K_{Sct}} + {K_{LoS}}} \right)}}} \right)^p}\notag\\
&\times \exp \left( { - \frac{{{\gamma _{th}}}}{{{\lambda _t}\left( {{K_{Sct}} + {K_{LoS}}} \right)}}} \right).
 \end{align}
\end{theorem}

\subsection{ERGODIC RATES}

There is a transformation from the OP expression to that of the ER \cite{Zhang_Semi_2024_Apr,Zhao_Ergodic_2022_Oct}. We denote the ER expression as $R_{er}$ and express it as:
\begin{align}\label{ER_def}
{R_{er}} = \mathbb{E}\left[\log_2\{1+\gamma_{SNR}\}\right]
=\frac{1}{{\ln 2}}\int_0^\infty  {\frac{{1 - {P_{out}}\left( {{\gamma _{th}}} \right)}}{{1 + {\gamma _{th}}}}} d{\gamma _{th}}.
\end{align}

Then, the following theorem derives the closed-form expression of the ER.

\begin{theorem}\label{T_ER}
By substituting the results of \textbf{\emph{Theorem} \ref{T_OP}} into \eqref{ER_def}, the ER expression for our SAGIN is derived as:
\begin{align}\label{ER_close}
{R_{er}} =& \frac{1}{{\ln 2}}\sum\limits_{k = 0}^{m - 1} {\binom{m-1}{k}\frac{{{{\left( {{K_{Sct}}} \right)}^{m - k - 1}}{{\left( {{K_{LoS}}} \right)}^k}}}{{{{\left( {{K_{Sct}} + {K_{LoS}}} \right)}^{m - 1}}}}} \notag\\
&\times \left( {{I_3} + \sum\limits_{p = 1}^k {\frac{{{{\left( {{K_{Sct}} + {K_{LoS}}} \right)}^p}}}{{p!{{\left( {{\lambda _t}} \right)}^p}}}} {I_4}} \right),
\end{align}
where we have:
\begin{subequations}
\begin{align}
  {R_{er}}\left| {_{p = 0}} \right. = & \frac{1}{{\ln 2}}\sum\limits_{k = 0}^{m - 1} {\binom{m-1}{k}\frac{{{{\left( {{K_{Sct}}} \right)}^{m - k - 1}}{{\left( {{K_{LoS}}} \right)}^k}}}{{{{\left( {{K_{Sct}} + {K_{LoS}}} \right)}^{m - 1}}}}} {I_3},\\
  {R_{er}}\left| {_{p \ge 1}} \right. =& \frac{1}{{\ln 2}}\sum\limits_{k = 0}^{m - 1} \binom{m-1}{k}\notag\\
  &\times \sum\limits_{p = 0}^k {\frac{{{{\left( {{K_{Sct}}} \right)}^{m - k - 1}}{{\left( {{K_{LoS}}} \right)}^k}}{I_4}}{{p!{{\left( {{\lambda _t}} \right)}^p}{{\left( {{K_{Sct}} + {K_{LoS}}} \right)}^{m - 1 - p}}}}} ,
\end{align}
\end{subequations}
with $I_3$ and $I_4$ formulated as:
\begin{subequations}
\begin{align}
   {I_3} =&  {-} \exp \left( {\frac{1}{{{\lambda _t}\left( {{K_{Sct}} {+} {K_{LoS}}} \right)}}} \right){\rm{Ei}}\left( { {-} \frac{1}{{{\lambda _t}\left( {{K_{Sct}} {+} {K_{LoS}}} \right)}}} \right),\\
   {I_4} =& \Gamma \left( {p + 1} \right)\Psi \left( {p + 1,p + 1;\frac{1}{{{\lambda _t}\left( {{K_{Sct}} + {K_{LoS}}} \right)}}} \right),
\end{align}
\end{subequations}
where the exponential integral is denoted as ${\rm{Ei}}\left( x \right) = \int_{ - \infty }^x {\frac{{\exp \left( t \right)}}{t}dt}$, $\Psi \left(  \cdot  \right)$ is the Tricomi confluent hypergeometric function \cite{table}, and $\Gamma(\cdot)$ is the complete Gamma function.
\begin{proof}
As for $p=0$, we exploit Eq. [2.3.4.3] of \cite{table} to derive $I_3$. As for $p \ge 1$, Eq. [2.3.6.9] of \cite{table} is harnessed to calculate $I_4$. The remaining straightforward derivations are omitted here due to the strict page limitation.
\end{proof}
\end{theorem}

\subsection{GOODPUT}
Finally, we calculate the GP expression, defined as:
\begin{align}\label{GP_def}
{R_{GP}} = \left(1-{\rm{BER}}\right) \times {R_{er}}.
\end{align}

\begin{theorem}\label{T_Goodputs}
Based on the results of \textbf{\emph{Theorem} \ref{T_BER}} and \textbf{\emph{Theorem} \ref{T_ER}}, we arrive at the final expression of the lower bound of GP, formulated as:
\begin{align}
{R_{GP}} \ge &\frac{1}{{\ln 2}}\left( {1 - \frac{1}{5}\exp \left( {\frac{{ - 3{\lambda _t}}}{{2\left( {M - 1} \right)}}} \right)} \right) \notag\\
& \times \sum\limits_{k = 0}^{m - 1} {\binom{m-1}{k}} \frac{{{{\left( {{K_{Sct}}} \right)}^{m - k - 1}}{{\left( {{K_{LoS}}} \right)}^k}}}{{{{\left( {{K_{Sct}} + {K_{LoS}}} \right)}^{m - 1}}}}\notag \\
& \times \left( {{I_3} + \sum\limits_{p = 1}^k {\frac{{{{\left( {{K_{Sct}} + {K_{LoS}}} \right)}^p}}}{{p!{{\left( {{\lambda _t}} \right)}^p}}}} {I_4}} \right).
\end{align}
\begin{proof}
This theorem is proved by substituting \eqref{BER_final} and \eqref{ER_close} into \eqref{GP_def}.
\end{proof}
\end{theorem}

\section{NUMERICAL RESULTS}

Our numerical results first validate the accuracy of the analytical derivations in terms of the following settings. The radius of the Earth $R$ is 6371.393 km. The altitude of the LEO $H$ is 300 km. The transmit power $P_s$ of the satellite is chosen from the range of $[30,50]$ dBm. As for the noise, we calculate it by $\sigma^2 = -170+10\times \log_{10}(BW)=-90$ dBm with $BW = 10^6$ Hz. As for the refractive index, we have $N_0 = 315$ and $h_0 =7.5$ km based on the ITU-R standard. The satellite transmits at a detected elevation angle of $\theta_0=60^{\circ}$ using a 2 GHz carrier. As for the space transmission, we set the path loss component as $\alpha=2$. We set the Shadowed-Rician distribution parameters to $m=4$, $K_{Sct} = 2\times b_0=0.2$, and $K = \Omega = 0.8$, thus $K_{LoS} = \frac{\Omega}{m}=0.2 $. The outage threshold $\gamma_{th}$ is 0.1 and the simulation number is $10^6$ times to calculate the average performance. Note that the units of ER and GP are bits per cell per second per Hertz.

\begin{figure}[!htb]
\centering
\includegraphics[width= 3.6in]{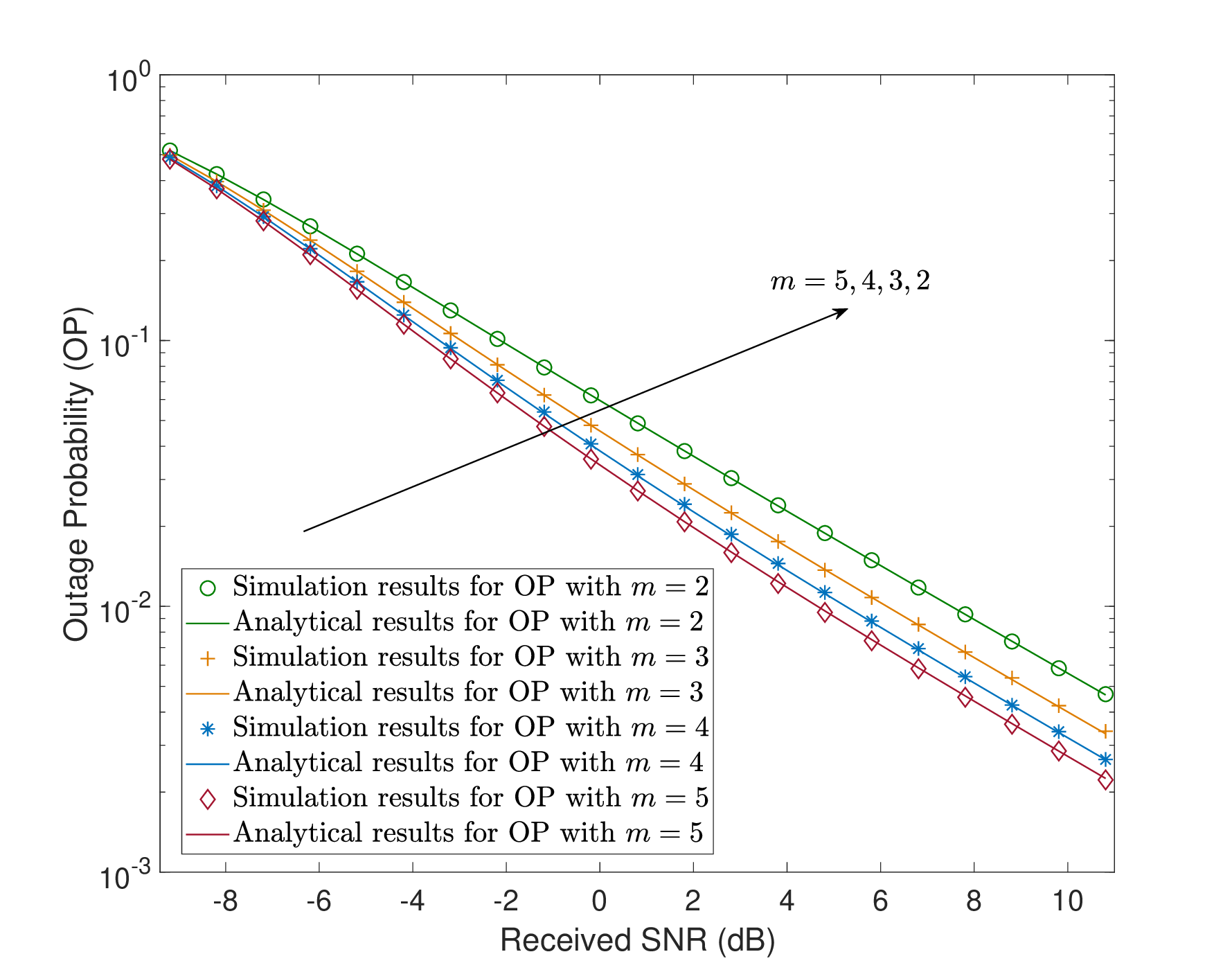}
\caption{The OP versus received SNR with different $m = [2,3,4,5]$ in the Shadowed-Rician fading channel (\textbf{\emph{Theorem} \ref{D_K}} and \textbf{\emph{Theorem} \ref{T_OP}}).}
\label{figure1}
\end{figure}

\begin{figure}[!htb]
\centering
\includegraphics[width= 3.6in]{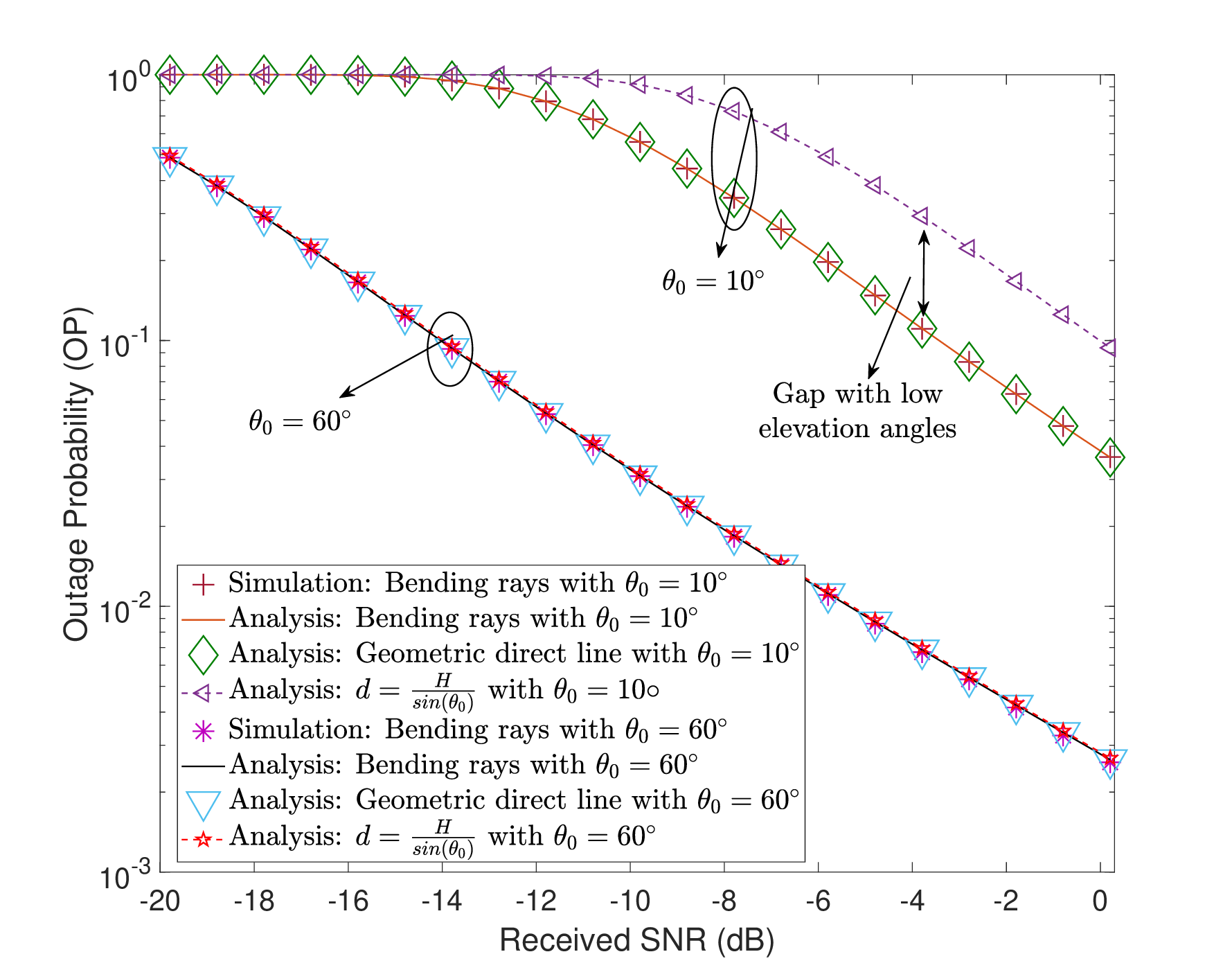}
\caption{The OP versus received SNR with different elevation angles: Bending rays, direct lines, and $\frac{H}{\sin{\theta_0}}$ (\textbf{\emph{Theorem} \ref{T_Bending}}, \textbf{\emph{Theorem} \ref{T_Bending2}}, \textbf{\emph{Theorem} \ref{T_straight}} and \textbf{\emph{Theorem} \ref{T_OP}}).}
\label{figure2}
\end{figure}

\begin{figure}[!htb]
\centering
\includegraphics[width= 3.6in]{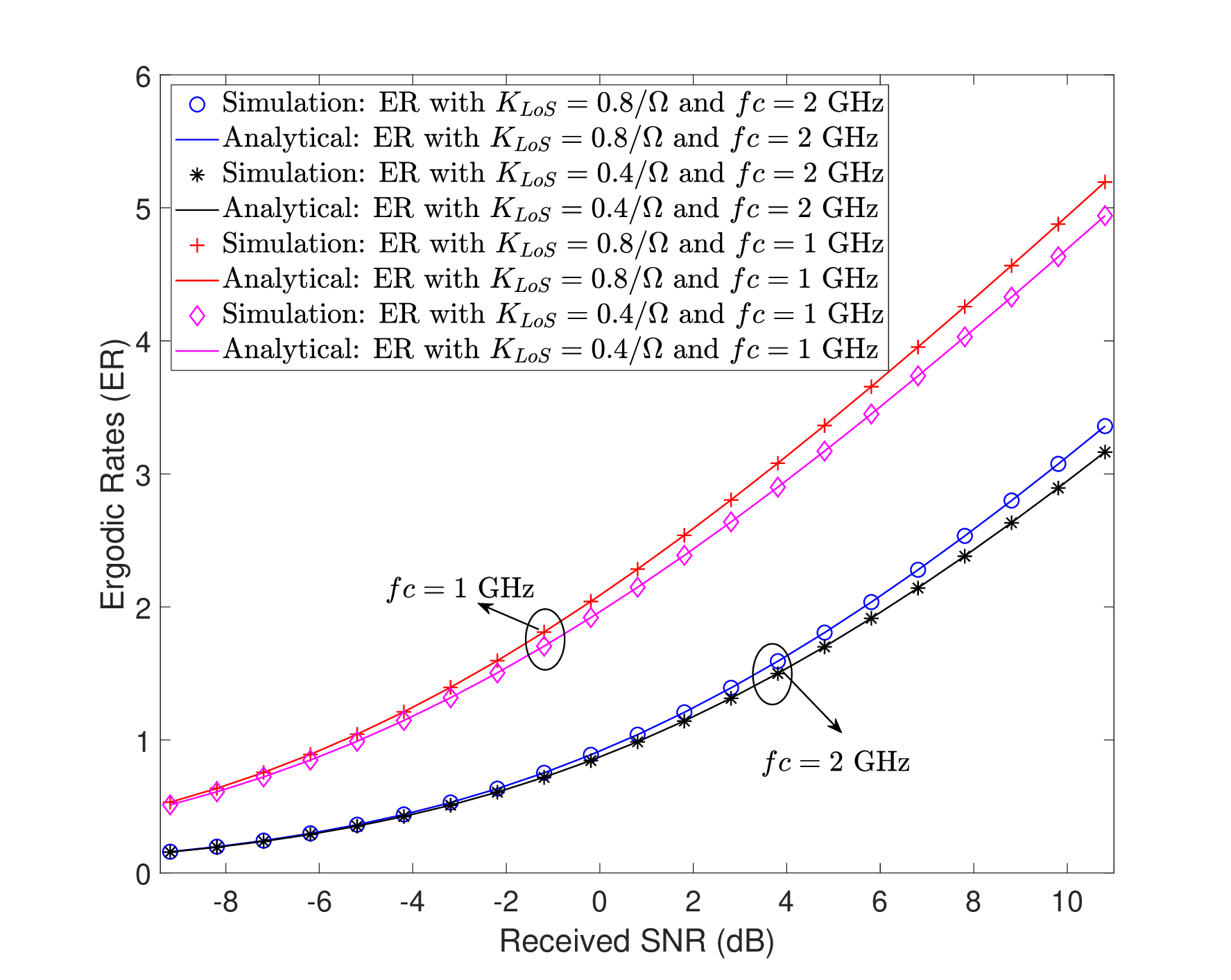}
\caption{The ER versus received SNR: the elevation of LoS components $K_{LoS}$ and carrier frequency $f_c$ (\textbf{\emph{Theorem} \ref{T_ER}}).}
\label{figure3}
\end{figure}

\begin{figure}[!htb]
\centering
\includegraphics[width= 3.6in]{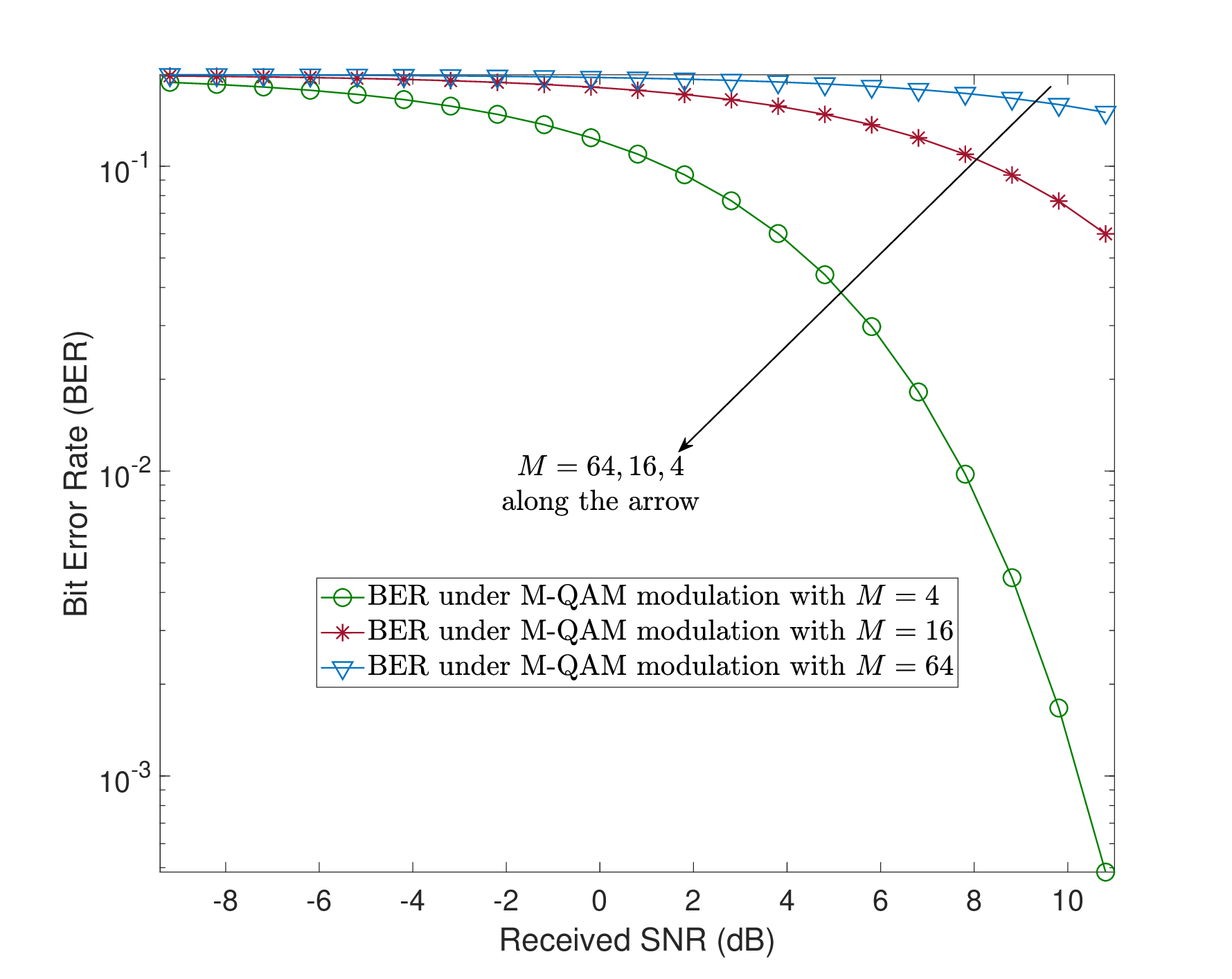}
\caption{The BER versus received SNR with 4-QAM, 16-QAM, and 64-QAM modulation (\textbf{\emph{Theorem} \ref{T_BER}}).}
\label{figure4}
\end{figure}

\begin{figure}[!htb]
\centering
\includegraphics[width= 3.6in]{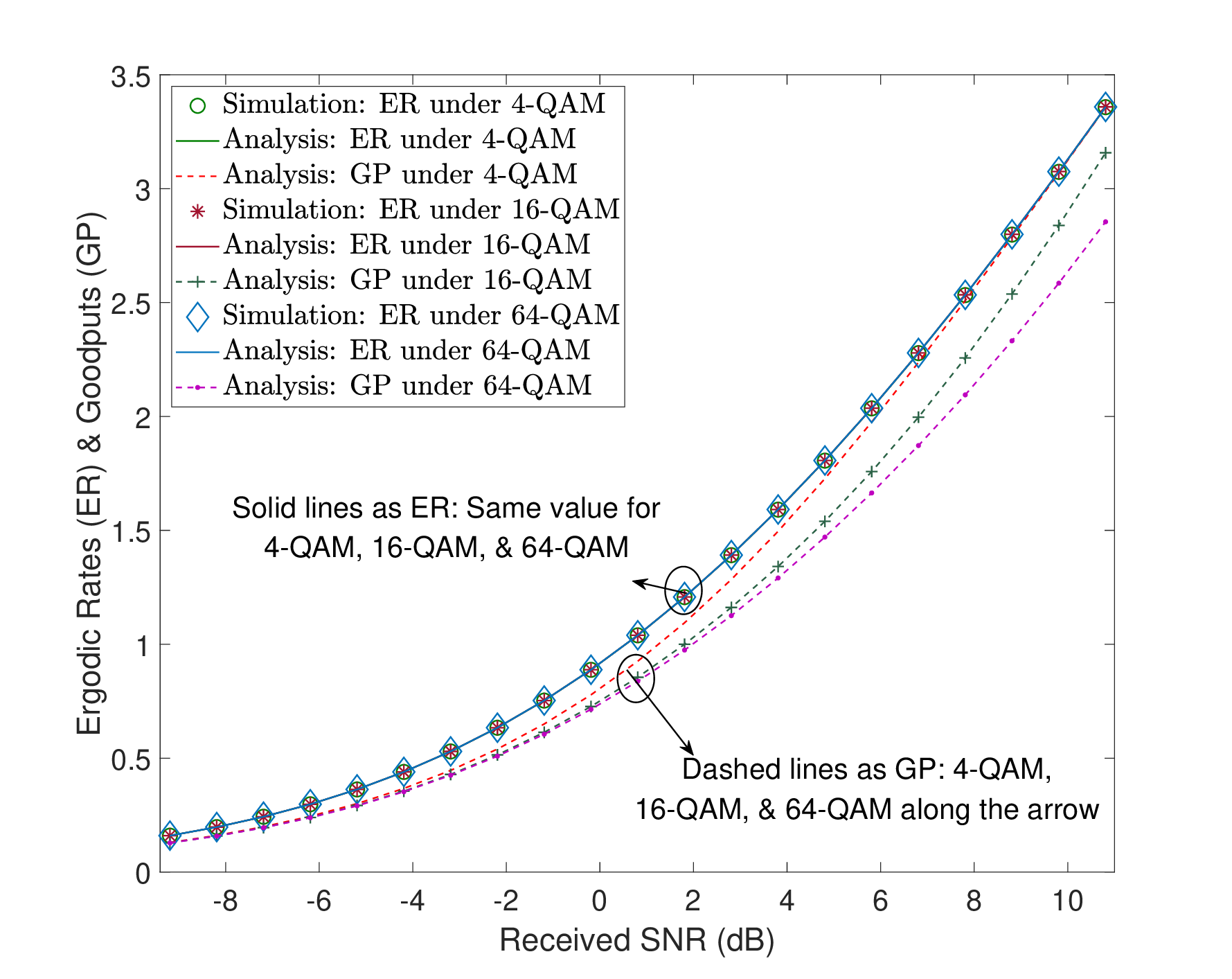}
\caption{A comparison between the ER and GP under different modulation methods (\textbf{\emph{Theorem} \ref{T_ER}}, and \textbf{\emph{Theorem} \ref{T_Goodputs}}).}
\label{figure5}
\end{figure}

In the simulation results, we assume that the Doppler effect has been well averaged out. Additionally, since we use a 2 GHz carrier, there is only slight molecular absorption in the L band, therefore we ignore it. Finally, to consider the best performance, the following simulation results are acquired under good weather conditions. The effects of rain, fog, and clouds are considered as a parameter affecting the following results.

In Fig. \ref{figure1}, the accuracy of the OP expression is validated versus the received SNR. Since the parameter $m$ in the Shadowed-Rician fading channels represents the influence of the shadowing effect on the LoS component, we compare the OP as expected of different values of $m$. The simulation results indicate that a higher $m$ has a lower OP. This is because increasing the parameter $m$ leads to reduced shadowing effects according to our model.

We evaluate the path loss for the different geometric models in Fig. \ref{figure2}, including bending rays considering the Earth's curvature, direct line considering the Earth's curvature, and the benchmark $d=\frac{H}{\sin{\theta_0}}$. In some papers, to reduce the complexity of the derivation, the Earth's surface is considered to be a horizontal plane, which deviates from the reality. Thus, according to our model, the distance used for path loss calculation is $d = H/\sin(\theta_0)$. As shown in the simulation results, the ``horizontal" assumption is acceptable for high elevation angles, while inaccuracies are inevitable when we omit the Earth's curvature to calculate the path loss for low elevation angles. Additionally, the refractivity effect is limited when we exploit the refractive index function in ITU-R for dry air and good weather conditions. However, this does not mean that the effect of refractivity remains low for other scenarios, especially for humid climates.

Fig. \ref{figure3} verifies the accuracy of the ER derivation. By considering different Rician factors $K$, the simulation results show that as expected, a higher value of $K$ maps to improved ER performance. This is because a higher $K$ represents stronger LoS components, resulting improved ER performance. We also conclude that high frequency bands degrade the performance of SAGINs. This is because the propagation paths in SAGINs are quite long, causing severe path loss. The path loss exponent is also higher when we have a higher carrier frequency, which further degrades the ER.

In Fig. \ref{figure4}, we evaluate the BER upper bound of different M-QAM schemes, which indicates that increasing $M$ increases the BER. Since the received SNR of SAGINs is typically low due to the severe path loss, the employment of 4-QAM is the most realistic. Finally in Fig. \ref{figure5}, we compare the ER and GP of different M-QAM schemes. The numerical results indicate that there is a gap between the ER and GP. This is because GP has only considered the error-free throughput. Hence, the GP has the potential to evaluate the performance when we consider different coding, modulation, and channel estimation induced errors.

\section{CONCLUSIONS}

A practical channel model was proposed for SAGINs, followed by investigating its long-term performance. The statistics of small-scale fading were analyzed to indicate the effect of the LoS and nLoS components. The path loss along the bending rays was calculated and the true geometry-based elevation angle was acquired. Furthermore, by considering the Earth's rotation, the normalized Doppler frequencies of both terrestrial and non-terrestrial users were characterized. This model also considered the effects of molecular absorption and weather conditions, like rain, fog, and clouds. By deriving the BER, OP, and ER, the upper bound of GP was calculated for SAGINs. The simulation results indicate that the bending rays slightly aggravate the path loss. As for small elevation angles, the geometric relationship, such as the Earth's curvature, cannot be ignored. The GP metric conveniently lends itself to evaluating different modulation methods, channel estimation errors, and Doppler effects.

\section*{APPENDIX~A: PROOF OF LEMMA \ref{Shadowed_Rician_channel}} \label{Appendix:A}
\renewcommand{\theequation}{A.\arabic{equation}}
\setcounter{equation}{0}

The confluent hypergeometric function of the first kind has a special-value form, which is expressed as:
\begin{align}
_1{F_1}\left( {m;1;{c_{ST}}x} \right) =& \underbrace {{L_{ - m}}\left( {{c_{ST}}x} \right)}_{\rm{Laguerre} \hspace*{0.2cm} \rm{polynomials}}\notag\\
 = &\exp \left( {{c_{ST}}x} \right){L_{m - 1}}( - {c_{ST}}),
\end{align}
based on the definition of Laguerre polynomials associated with negative values, and denoted as ${L_{ - n}}(x) = {e^x}{L_{n - 1}}( - x)$.

By further substituting the series expression of Laguerre polynomials, given by ${L_n}(x) = \sum\limits_{k = 0}^n \binom{n}{k} \frac{{{{( - 1)}^k}}}{{k!}}{x^k}$, into the PDF of the Shadowed-Rician channel, we arrive at the final expression of the PDF:
\begin{align}
{f_{{{\left| {{h_{ST}}} \right|}^2}}}(x) = {a_{ST}}\sum\limits_{k = 0}^{m - 1} {\binom{m-1}{k}\frac{{{{\left( {{c_{ST}}} \right)}^k}}}{{k!}}{x^k}\exp \left( { - {e_{ST}}x} \right)}.
\end{align}

By exploiting the definition of the lower incomplete Gamma function, we obtain the associated CDF expression as:
\begin{align}
& {F_{{{\left| {{h_{ST}}} \right|}^2}}}(y) \notag\\
& ={a_{ST}}\sum\limits_{k = 0}^{m - 1} {\binom{m-1}{k}\frac{{{{\left( {{c_{ST}}} \right)}^k}}}{{k!}}\int_0^y {{x^k}} \exp \left( { - {e_{ST}}x} \right)dx} \notag\\
 &= {a_{ST}}\sum\limits_{k = 0}^{m - 1} {\binom{m-1}{k}\frac{{{{\left( {{c_{ST}}} \right)}^k}}}{{k!e_{ST}^{k + 1}}}\underbrace {\int_0^{{e_{ST}}y} {{t^{\left( {k + 1} \right) - 1}}} \exp \left( { - t} \right)dt}_{{\rm{Incomplete\hspace*{0.1cm} Gamma \hspace*{0.1cm} Function}}}}\notag \\
 &= {a_{ST}}\sum\limits_{k = 0}^{m - 1} {\binom{m-1}{k}\frac{{{{\left( {{c_{ST}}} \right)}^k}}}{{k!e_{ST}^{k + 1}}}\gamma \left( {k + 1,{e_{ST}}y} \right)}.
\end{align}

Note that $\Gamma(\cdot)$ is the complete Gamma function. Given a positive integer $a$, the lower incomplete Gamma function can be expressed by the summation of multiple exponential functions \cite{Zhang_Semi_2023_Apr,Yang_Performance_2016_Jan,Kang_Capacity_2006_Jan}, denoted as:
\begin{align}
\gamma (a,b) &= \Gamma (a) - \Gamma (a,b)= \Gamma (a) - \sum\limits_{p = 0}^{a - 1} {\frac{{(a - 1)!}}{{p!}}} \exp ( - b){b^p}.
\end{align}

With the aid of the above series expansion of the complete Gamma function, the CDF expression is further expressed as:
\begin{align}\label{A5}
{F_{{{\left| {{h_{ST}}} \right|}^2}}}(y) = &\underbrace {\sum\limits_{k = 0}^{m - 1} {\binom{m-1}{k}\varsigma (k)} }_{{I_1}} - \sum\limits_{k = 0}^{m - 1} \binom{m-1}{k}\varsigma (k)\notag\\
&\times \sum\limits_{p = 0}^k {\frac{{{{\left( {{e_{ST}}y} \right)}^p}}}{{p!}}} \exp \left( { - {e_{ST}}y} \right).
\end{align}

By exploiting the binomial expansion, the expression of $I_1$ in \eqref{A5} is proved to be equal to one, which is shown as follows:
\begin{align}
{I_1} &= {a_{ST}}\sum\limits_{k = 0}^{m - 1} {\binom{m-1}{k}\frac{{{{\left( {{c_{ST}}} \right)}^k}}}{{e_{ST}^{k + 1}}}} \notag\\
 & = \frac{{\sum\limits_{k = 0}^{m - 1} {\binom{m-1}{k}} {{\left( {2{b_0}m} \right)}^{m - k - 1}}{\Omega ^k}}}{{{{\left( {2{b_0}m + \Omega } \right)}^{m - 1}}}}\notag\\
& = \frac{{{{\left( {2{b_0}m + \Omega } \right)}^{m - 1}}}}{{{{\left( {2{b_0}m + \Omega } \right)}^{m - 1}}}} = 1.
\end{align}

Consequently, the final CDF expression is arrived at and the proof ends.

\renewcommand{\theequation}{C.\arabic{equation}}
\setcounter{equation}{4}
\begin{figure*}[!hb]
\hrulefill
\begin{align}\label{Doppler1}
\frac{{\Delta f}}{{{f_c}}} =&  - \frac{{\dot s(t)}}{c} =  - \frac{1}{c}\frac{{R{H_{os}}\cos \gamma \left( {{t_0}} \right)\sin \left( {\psi \left( t \right) - \psi \left( {{t_0}} \right)} \right)\dot \psi \left( t \right)}}{{\sqrt {{R^2} + H_{os}^2 - 2R{H_{os}}\cos \left( {\psi \left( t \right) - \psi \left( {{t_0}} \right)} \right)\cos \gamma \left( {{t_0}} \right)} }}\notag\\
 = & - \frac{1}{c}\frac{{R{H_{os}}\sin \left( {\psi \left( {t,{t_0}} \right)} \right)\cos \left( {{{\cos }^{ - 1}}\left( {\frac{{R\cos {\theta _{\max }}}}{{{H_{os}}}}} \right) - {\theta _{\max }}} \right){{\dot \psi \left( t \right)}}}}{{\sqrt {{R^2} + H_{os}^2 - 2R{H_{os}}\cos \left( {\psi \left( {t,{t_0}} \right)} \right)\cos \left( {{{\cos }^{ - 1}}\left( {\frac{{R\cos {\theta _{\max }}}}{{{H_{os}}}}} \right) - {\theta _{\max }}} \right)} }}.
\end{align}
\end{figure*}

\section*{APPENDIX~B: PROOF OF THEOREM \ref{T_Bending2}} \label{Appendix:B}
\renewcommand{\theequation}{B.\arabic{equation}}
\setcounter{equation}{0}

Based on the ray-tracing method, the terrestrial range is given by the following integral as:
\begin{align}
G &= \int_0^H {\frac{1}{{\left( {1 + \frac{h}{R}} \right)\sqrt {{{\left[ {\frac{{n\left( {1 + \frac{h}{R}} \right)}}{{{n_0}\cos \left( {{\theta _0}} \right)}}} \right]}^2} - 1} }}dh}.
\end{align}

Let us define the refractivity by a general exponential function as $n = 1 + {\rho _0}\exp \left( { - kh} \right)$. Upon replacing ${\cos ^2}\left( {{\theta _0}} \right)$ by the equivalent formula of $1 - {\sin ^2}\left( {{\theta _0}} \right)$, the expression of the ground range is further expressed as:
\begin{align}\label{G_computation}
G = \int_0^H {\frac{{\left( {1 + {\rho _0}} \right)\cos \left( {{\theta _0}} \right)dh}}{{\left( {1 + \frac{h}{R}} \right)\sqrt {\mu  + \upsilon \left( h \right) + \omega \left( h \right) + \upsilon \left( h \right)\omega \left( h \right)} }}} ,
\end{align}
where the parameters in \eqref{G_computation} are reformulated as:
\begin{subequations}
\begin{align}
&\mu  = {\left( {1 + {\rho _0}} \right)^2}{\sin ^2}{\theta _0} - 2{\rho _0} - \rho _0^2 ,\\
&\upsilon \left( h\right) = 2{\rho _0}\exp \left( { - kh} \right) + \rho _0^2\exp \left( { - 2kh} \right) ,\\
&\omega \left( h\right)  = \frac{{2h}}{R} + \frac{{{h^2}}}{{{R^2}}}.
\end{align}
\end{subequations}

Subsequently, the length of the bending rays is further expressed as:
\begin{align}\label{d_rf_int}
{d_{rf}}\left( h \right) = \int_0^H {\frac{{{n^2}\left( h \right)\left( {1 + \frac{h}{R}} \right)dh}}{{\sqrt {\mu  + \upsilon \left( h \right) + \omega \left( h \right) + \upsilon \left( h \right) \omega \left( h \right)} }}}.
\end{align}

With the aid of the Chebyshev-Gauss quadrature, the closed-form expression in \eqref{G_computation} is expressed as \eqref{G_computation2}. Following the same derivation processes, the closed-form expression of \eqref{d_rf_int} is formulated in \eqref{d_rf_2}.

\section*{APPENDIX~C: PROOF OF THEOREM \ref{T_Doppler}} \label{Appendix:C}
\renewcommand{\theequation}{C.\arabic{equation}}
\setcounter{equation}{0}

\begin{figure}[!htb]
\centering          
	\includegraphics[width= 3.3in]{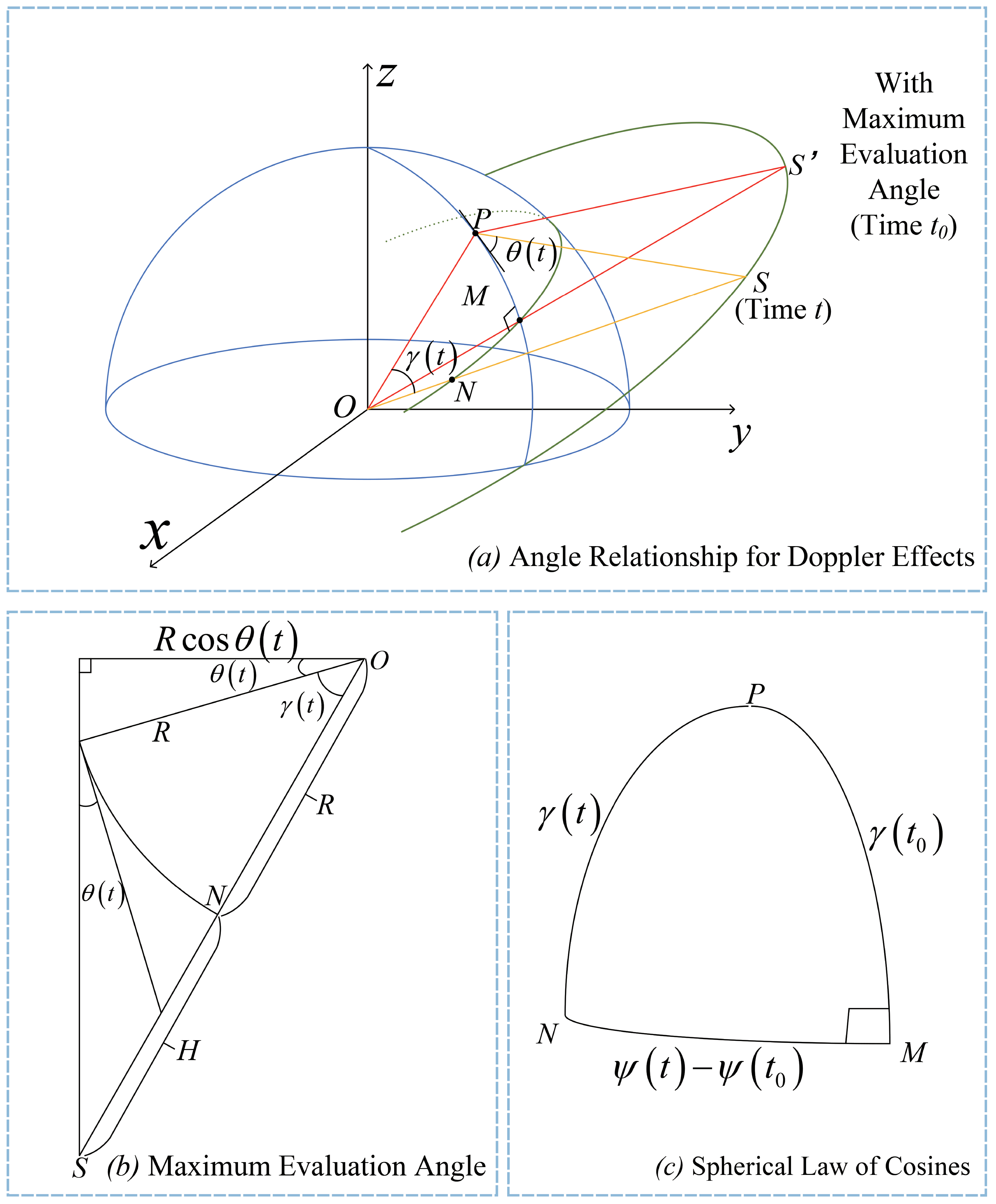}   
    \caption{The notation of angles from the satellite to the ground user for \textbf{\emph{Theorem} \ref{T_Doppler}}.}
        \label{Doppler_fig}
\end{figure}

The geometric relationship between the position of the satellite and that of the user is given in Fig. \ref{Doppler_fig}. Given the maximum elevation angle $\theta_{max}$, the slant range $s(t)$ is given by the law of cosines, formulated as:

\begin{align}\label{slant}
s(t) = \sqrt {{R^2} + H_{os}^2 - 2R{H_{os}}\cos \gamma \left( t \right)},
\end{align}
where ${\gamma \left( t \right)}$ is the angle at time $t$ generated by the origin of the Earth, the position of the satellite at time instant $t_0$, and that at time $t$.

The spherical law of cosines is exploited, which is expressed as $\cos c = \cos a \cos b + \sin a \sin b \cos C$. Since the angle $\angle PMN$ equals $\pi/2$, we have an angular relationship expressed as:
\begin{align}\label{spherical}
\cos \gamma \left( t \right) = \cos \left( {\psi \left( t \right) - \psi \left( {{t_0}} \right)} \right)\cos \gamma \left( {{t_0}} \right),
\end{align}
where ${\psi \left( t \right) - \psi \left( {{t_0}} \right)}$, $\gamma \left( t_0\right)$, and $\gamma \left( t\right)$ are the angles constituted by the origin of the Earth and the related arc along the surface of the Earth, which is shown in Fig. \ref{Doppler_fig}. \emph{(c)}.

By substituting \eqref{spherical} into \eqref{slant} and differentiating \eqref{slant} with respect to time $t$, we have:
\begin{align}\label{slant_dt}
\dot s(t) =& \frac{{ds(t)}}{{dt}} \notag\\
=& \frac{{R{H_{os}}\cos \gamma \left( {{t_0}} \right)\sin \left( {\psi \left( t \right) - \psi \left( {{t_0}} \right)} \right)\frac{{d\psi \left( t \right)}}{{dt}}}}{{\sqrt {{R^2} + H_{os}^2 - 2R{H_{os}}\cos \left( {\psi \left( t \right) - \psi \left( {{t_0}} \right)} \right)\cos \gamma \left( {{t_0}} \right)} }}\notag\\
 =& \frac{{R{H_{os}}\cos \gamma \left( {{t_0}} \right)\sin \left( {\psi \left( t \right) - \psi \left( {{t_0}} \right)} \right)\dot \psi \left( t \right)}}{{\sqrt {{R^2} + H_{os}^2 - 2R{H_{os}}\cos \left( {\psi \left( t \right) - \psi \left( {{t_0}} \right)} \right)\cos \gamma \left( {{t_0}} \right)} }},
\end{align}
where we define the differential equation versus time as $\dot f(t) = df(t)/dt$.

We assume that at the time instant $t_0$, the position of the satellite is at the point of the maximum elevation angle $\theta_{max}$, denoted as $\theta_0 = \theta_{max}$. Hence, based on the associated triangle seen in Fig. \ref{Doppler_fig}. \emph{(b)}, the angle relationship satisfies:
\begin{align}\label{angle_relationship2}
\cos \left( {{\theta _{\max }} + \gamma \left( {{t_0}} \right)} \right) = \frac{{R\cos {\theta _{\max }}}}{{{H_{os}}}},
\end{align}
where $H_{os} = R+H$ represents the straight distance from the origin of the Earth to the satellite.

By substituting \eqref{angle_relationship2} into \eqref{slant_dt}, we arrive at the normalized Doppler frequency in terms of the maximum elevation and time $t$ as seen in \eqref{Doppler1}, where we have $\psi \left( {t,{t_0}} \right) = \psi (t) - \psi \left( {{t_0}} \right)$.

By considering the relative angular velocity shown as \eqref{a_v} in the Section II, we have the final expression in this theorem.

\section*{APPENDIX~D: PROOF OF LEMMA \ref{L_E_h_ST}} \label{Appendix:E}
\renewcommand{\theequation}{D.\arabic{equation}}
\setcounter{equation}{0}

The expectation of ${{{\left| {{h_{ST}}} \right|}^2}}$ is expressed as:
\begin{align}\label{E_def_h}
{\mathbb{E}}\left[ {{{\left| {{h_{ST}}} \right|}^2}} \right] = \int_0^\infty  x {f_{{{\left| {{h_{ST}}} \right|}^2}}}(x)dx.
\end{align}

By substituting \eqref{PDF_h_ST} into \eqref{E_def_h} and exploiting Eq. [2.3.4.1] of \cite{table}, we have:
\begin{align}\label{E_h_2}
\hspace*{-0.2cm}\mathbb{E}\left[ {{{\left| {{h_{ST}}} \right|}^2}} \right] {=} &  {a_{ST}}\sum\limits_{k {=} 0}^{m { -} 1} {\binom{m {-} 1}{k}\frac{{{{\left( {{c_{ST}}} \right)}^k}}}{{k!}}} \int_0^\infty  {\frac{{\exp \left( { { -} {e_{ST}}x} \right)}}{{{x^{ - k + 1}}}}} dx \notag\\
{=} & {a_{ST}}\sum\limits_{k = 0}^{m {-} 1} {\binom{m-1}{k}\left( {k + 1} \right)\frac{{{{\left( {{c_{ST}}} \right)}^k}}}{{{{\left( {{e_{ST}}} \right)}^{\left( {k + 2} \right)}}}}} .
\end{align}

By substituting the parameters $a_{ST}$, $c_{ST}$, and $e_{ST}$ into \eqref{E_h_2}, the targeted expectation is further derived as:
\begin{align}\label{E_I2}
\mathbb{E}\left[ {{{\left| {{h_{ST}}} \right|}^2}} \right] = \frac{{2{b_0}\underbrace {\sum\limits_{k = 0}^{m - 1} {\binom{m-1}{k}{{\left( {\frac{\Omega }{{2m{b_0}}}} \right)}^k}} \left( {k + 1} \right)}_{{I_2}}}}{{{{\left( {1 + \frac{\Omega }{{2{b_0}m}}} \right)}^{m - 2}}}}.
\end{align}

We then arrive at a specific form, which is expressed as the powers of a binomial, denoted as $1+ \frac{\Omega}{2 b_0 m}$, by the Binomial theorem. In the following, we express $I_2$ of \eqref{E_I2} as:
\begin{align}\label{I_2}
{I_2} = {\left( {1 {+} \frac{\Omega }{{2m{b_0}}}} \right)^{m - 1}} {+} \left( {m - 1} \right)\left( {\frac{\Omega }{{2m{b_0}}}} \right){\left( {1 {+} \frac{\Omega }{{2m{b_0}}}} \right)^{m - 2}}.
\end{align}

Finally, we substitute \eqref{I_2} into \eqref{E_I2} to obtain the final expectation expression.

\bibliographystyle{IEEEtran}
\bibliography{mybib}

\end{document}